\documentclass[english,a4paper,11pt]{scrartcl}
\usepackage[latin1]{inputenc}
\usepackage[dvips]{graphics}
\usepackage[dvips]{graphicx}
\usepackage{amssymb,amsmath}     
\usepackage{cite}

\newcommand{\bc}{\begin{center}}
\newcommand{\ec}{\end{center}}
\newcommand{\bq}{\begin{quote}}
\newcommand{\eq}{\end{quote}}
\newcommand{\bi}{\begin{itemize}}
\newcommand{\ei}{\end{itemize}}
\newcommand{\bn}{\begin{enumerate}}
\newcommand{\en}{\end{enumerate}}
\newcommand{\bd}{\begin{description}}
\newcommand{\ed}{\end{description}}
\newcommand{\be}{\begin{eqnarray}}
\newcommand{\ee}{\end{eqnarray}}
\newcommand{\beq}{\begin{equation}}
\newcommand{\eeq}{\end{equation}}
\newcommand{\bs}{\begin{split}}
\newcommand{\es}{\end{split}}
\newcommand{\bdf}{\begin{definition}}
\newcommand{\edf}{\end{definition}}
\newcommand{\bt}{\begin{tabbing}}
\newcommand{\et}{\end{tabbing}}
\newcommand{\bfig}{\begin{figure}}
\newcommand{\efig}{\end{figure}}

\newcommand{\refsec}[1]{(Sec.~\ref{#1})}
\newcommand{\refeq}[1]{(Eq.~\ref{#1})}

\newcommand{\LT}{LT$\alpha_1\beta_2$}

\newcommand{\MICRON}{$\mu\rm m$}

\begin{document}
\title{Mechanisms of organogenesis of primary lymphoid follicles}
\date{\today}
\author{Tilo Beyer and Michael Meyer-Hermann\\
Frankfurt Institute for Advanced Studies,
Johann Wolfgang Goethe-University,\\
Max-von-Laue-Str. 1,
60438 Frankfurt Main, Germany\\[0.2cm]
Corresponding author: Tilo Beyer, E-mail: {\tt tbeyer@fias.uni-frankfurt.de}}

\maketitle

\begin{abstract}
Primary lymphoid follicles in secondary lymphoid tissue of mammals are the
backbone for the formation of follicular dendritic cell networks. These are
important for germinal center reactions.
In the context of organogenesis molecular requirements for the formation
of follicles have been identified.
The present study complements this work
with a simulation of the dynamics of the primary lymphoid follicle formation. 
In contrast to other problems of pattern formation,
here, the homeostasis of the cell population
is not governed by a growth-death balance 
but by a flow equilibrium of migrating cells. The influx
of cells into secondary lymphoid tissue was 
extensively studied while 
less information is available about the efflux 
of lymphocytes from secondary lymphoid tissues.
This study formulates the minimal requirements 
for cell efflux that guarantee a flow equilibrium and,
thus, a stable primary lymphoid follicle.
The model predicts that in addition to already 
identified mechanisms a negative regulation of the
generation of follicular dendritic cells is required.
Furthermore, 
a comparison with data concerning the microanatomy
of secondary lymphoid tissues yields
the conclusion that dynamical changes
during the formation of FDC networks
of the lymphatic endothelium are necessary
to understand the genesis and maintenance of follicles.
\end{abstract}
{\small
{\em Keywords:
lymphocytes, follicular dendritic cells, chemotaxis, spleen and lymph nodes, negative regulation}\\[0.3cm]
List of abbreviations:\\
{\bf ELV} efferent lymphatic vessel,
{\bf FDC} follicular dendritic cell,
{\bf FRC} fibroblastic reticular cell,
{\bf HEV} high endothelial venule,
{\bf LE}  lymphatic endothelium,
{\bf LN}  lymph node,
{\bf LT}  lymphotoxin,
{\bf PLF} primary lymphoid follicle,
{\bf S1P} sphingosine 1-phosphate,
{\bf S1P$_1$} S1P receptor 1,
{\bf SLT} secondary lymphoid tissue,
{\bf TNF} tumor necrosis factor
}
\newpage
\section{Introduction}
The primary lymphoid follicle (PLF) is the precursor state 
of germinal centers in secondary
lymphoid tissue (SLT) \cite{Maclennan:1994}. 
The most important constituents are naive B cells 
and follicular dendritic cells (FDC).
PLF serve as a filter for antigen 
and bring B cells into contact with antigen presented by FDC
in order to start a germinal center reaction \cite{Kosco-vilbois:2003}. 

In mammals PLF first develop 
around birth
\cite{Nicander:1991,Griebel:1996,Makala:2002,Mebius:1997,Nishikawa:2003,Finke:2005}.
During the past decade many molecules relevant to the 
formation of PLF could be isolated.
With an extensive number of knockout and antibody-blocking 
experiments the contributions
of TNF-$\alpha$, LT$\alpha_1\beta_2$, LT$\alpha_3$, LIGHT, and TRANCE
secondary lymphoid organ structure and in the formation of PLF
have been identified
(reviewed in \cite{Fu:1999,Muller:2003,Tumanov:2003}).

The homeostatic chemokines CCL19, CCL21 and CXCL13 
are involved in guiding the lymphocytes to their
compartment (reviewed in \cite{Cyster:2005}).
CXCL13 is secreted by FDC and acts as chemoattractant 
for B cells via the CXCR5 chemokine receptor.
CCL19 and CCL21 are produced by different cell populations in the T zone of SLT.
The common receptor CCR7 is expressed by T cells.
The expression of these chemokines and the presence of FDC depend
on LT$\alpha_1\beta_2$, TNF-$\alpha$ and related
molecules. 
The precise molecular interactions and chemokine
expression profiles slightly vary for different
SLT \cite{Fu:1999,Muller:2003,Tumanov:2003},
however, in most SLT a persistent LT$\alpha_1\beta_2$-stimulus is needed for both
the induction and the maintenance of FDC networks
\cite{Mackay:1998}.
Further studies suggested that the maintenance of the PLF structure is mediated by 
a positive feedback loop \cite{Ansel:2000}:
B cells are stimulated by CXCL13 to express high levels of LT$\alpha_1\beta_2$. 
This in turn stimulates FDC to produce CXCL13.

An interesting property of PLF is their equal size of several hundred $\mu\rm m$~in
all mammals ranging from mice to horses
\cite{Garside:1998,Bhalla:1981,Ohtani:2003,Halleraker:1994,Brachtel:1996,Kasajima-akatsuka:2006,Belz:1998,Kumar:2006}.
Thus, the follicle size does not scale with the size of the animals or
of the organs under consideration, which is considered to be a strong
boundary condition for simulations of PLF formation and maintenance.
The formation of PLF is studied with a simulation 
based on a previously introduced  agent-based model \cite{beyer:2006a} 
on top of a regular triangulation
\cite{Schaller:2004,Schaller:2005,beyer:2005}. A clear separation of B and T cells
is achieved using chemoattraction
to an exit spot for the lymphocytes overlaying the different chemokine responses of B and T cells
to their homing chemokines CXCL13, CCL21, and CCL19.
Formation of a PLF and an adjacent T zone is found
when assuming the generation of FDC from T zone stromal cells 
by sufficiently large B cell aggregates.
However, then, the location of the PLF relative to
lymphatic vessels is not in agreement with experiment.
The present study concentrates on the correct formation of the lymphatic vessels
and the PLF without considering the dynamics of the T zone of T cells.
The model includes lymphangiogenesis,
which dramatically changes the dynamics of the PLF formation, its geometry 
and also the regulation mechanisms involved in PLF maintenance 
and size regulation.
It is found that a negative regulation of the FDC generating process
is required.
Furthermore it is demonstrated that a chemotactic activity
of B cells for S1P is in contradiction to PLF morphology,
suggesting that S1P chemotaxis is not active during
PLF formation and maintenance {\em in vivo},
which was also found in experiment \cite{Wei:2005}.
\section{Model of primary lymphoid follicle formation}
This section describes the properties of a minimal model 
for PLF formation.
In addition the model of lymphangiogenesis is discussed. 
The mathematical framework is briefly summarized
in the method section \refsec{sec-method} and is
presented in more detail in \cite{beyer:2006a}.

\subsection{Flux of naive B cells}\label{sec-flux}
Essential ingredients to study the formation of the PLF are the B cell flux, the FDC and the FDC
precursor cells. The B cells are constantly entering SLT via high endothelial venules (HEV)
in the cases of mucosa-associated lymphoid tissue
(MALT) and lymph nodes (LN) or along the central arterioles through the peri-arteriolar sheath
of the spleen (reviewed in \cite{Cyster:2005}).
The exit route of lymphocytes through lymphatic endothelium (LE) is less clearly identified. 
Recent experimental data (see \cite{Cyster:2005}) suggest
that B cells leave MALT via efferent lymphatic vessels (ELV).
In LN and spleen the lymphatic sinuses guide lymphocytes to draining ELV.
As the sinuses merge with the ELV
they are considered to belong to the efferent lymphatic system
in the simulation.

The morphology of the LE 
is very similar across the different SLT and species.
Studies using corrosion casts revealed that the LE is formed around the follicles.
The precise shape ranges from half-open baskets around the follicle base 
to almost closed shells enclosing the whole
follicle. The LE forms a dense network in the non-follicular area 
\cite{Belz:1998,Drinker:1933,Ewijk:1980,Ohtani:1986,Ohtani:1991,Belisle:1990,Pellas:1990,Belz:1995,Azzali:2000,Azzali:2002,Azzali:2003}.

The role of LE in lymphocyte emigration is supported by a study 
of the corresponding structure in the Bursa of chicken
 \cite{Ekino:1979} and the identification of openings in 
the LE nearby follicles \cite{Drinker:1933,Ohtani:1991}.
The exit of lymphocytes via LE is further supported by
a newly identified mannose receptor on LE. This receptor may guide the specific
exit route of lymphocytes via the lymphatic sinuses \cite{Irjala:2001}. It
is almost exclusively found on the lymphatic vessels 
in the LN and binds CD62L, thus, guiding lymphocytes to the ELV and
allowing their exit from LN. 

S1P might be involved in the regulation of lymphocyte egress from SLT
\cite{Cyster:2005,Matloubian:2004,Sallusto:2004,Lo:2005}. 
S1P acts as a chemoattractant for
lymphocytes {\em in vitro} \cite{Matloubian:2004}.
Thus, the dynamics of the S1P receptor S1P$_1$ is considered in the simulation:
The LE is widely engulfing follicles such that lymphocytes have to migrate 
around or across these structures in order to reach the follicles.
Therefore a minimal transit time through SLT
should be governed by some mechanism that prevents
the lymphocytes from entering and transmigrating across the LE. 
In the simulation it is assumed that S1P$_1$ is downregulated 
when B cell enter SLT and required for
lymphocyte emigration. This is supported 
by studies using the S1P$_1$ agonist FTY720
\cite{Matloubian:2004,Sallusto:2004,Lo:2005,Cyster:2005}.
There is good agreement between the minimal
transit time of lymphocytes \cite{Pellas:1990,Pellas:1990b,Young:1999,Srikusalanukul:2002}
and the time required to fully upregulate S1P$_1$ on lymphocytes \cite{Lo:2005}
both of which are in the order of 3 hours.

\subsection{Origin of FDC}
One fundamental question in the PLF formation is: Where do the FDC come from?
The most commonly accepted view is
that FDC are derived from stromal cells which might be
related to the stromal cells observed in the T zone.
These stromal cells are fibroblastic reticular cells (FRC) 
\cite{Crivellato:1997,Kaldjian:2001,Nolte:2003,Balogh:2004}.
Comparative reviews supporting the mesenchymal
origin of FDC can be found in \cite{Nierop:2002}.
The evidence for the relation of FDC to FRC is given by 
shared markers \cite{Bofill:2000}. There exists
a gradual decrease of the fibroblast marker ASO-2 from 
the T zone to the heart of a germinal center.
This decrease is accompanied by a gradual morphological change 
from the 'classical fibroblastic' to the typical FDC morphology 
\cite{Szakal:1985}. A similar change of marker expression
has been observed in the PLF although a few intermediate steps are missing.
Indirect evidence for the reverse transition from FDC to stromal cells
is provided by
culture experiments in which purified FDC gradually loose characteristic
FDC marker one by one, finally turning stromal cell-like again
\cite{Clark:1992,Chang:2003,Tsunoda:1997}. This suggests
that persistent stimuli are required to maintain the FDC phenotype.
Thus, the study of typical FDC marker expression indicates a smooth
transition from stromal cell precursors to FDC by subsequent accumulation
of 'FDC-ness' of the stromal cells.
A recent study of FDC development in human LN 
supports the idea of local FDC differentiation:
FDC seem to acquire more and more markers like those found in culture experiments
extending the 'FDC-ness' picture to several other markers \cite{Kasajima-akatsuka:2006}).

Observations of slow replacement of FDC by 
migratory precursors and of inefficient seeding by bone
marrow-derived cells argue against a hematopoietic origin 
of FDC \cite{Cyster:2000}. 
However, a clear relationship between FRC and FDC could not be identified in a study
based on newly developed markers \cite{Balogh:2004}.
Bone marrow transfer
into severe combined immunodeficiency (SCID)
mice revealed donor-derived FDC suggesting that the stromal cells may originate
from the bone marrow \cite{Kapasi:1998}.
TNFR-I knockouts and mice given injections of a blocking TNFR-I-Ig
exhibit immature FDC or FDC precursors around splenic follicle suggesting migrative FDC progenitors
\cite{Kuprash:2005}. Similar conclusions have been drawn from TNF-$\alpha^{-/-}$~and TNFR-I$^{-/-}$ \cite{Pasparakis:2000}.
More support for the immigration of FDC precursor instead of local differentiation comes from the expression pattern
of TuJ1, a microtubule-protein \cite{Lee:2005}. The broad distribution with single positive cells scattered around
lymphoid follicles suggests migratory activity of FDC.
Reconciling both theories, FDC might be a heterogeneous population that can develop from hematopoietic and stromal
precursors \cite{Crivellato:2004,Cyster:2000}. This favor the vantage point adopted in this study considering
FDC as a state of cells rather than a differentiated form of some cell type.

In the simulation FDC differentiate from FRC.
This can be viewed as FDC being an 'excited' state of stromal cells.
However, this must not be interpreted in a strict sense.
The mechanism in the simulation is indistinguishable of the interpretation
that FRC and FDC have a common
progenitor which can develop in either FRC or FDC depending on the external stimuli.
Newly developing FDC replace existing FRC at their site of generation.

In a simplified PLF model it could be shown that sufficiently large aggregates of B cells
forming around stromal cells can induce the presence of FDC if basal levels of LT$\alpha_1\beta_2$ are expressed
on the B cells \cite{beyer:2006a}. The migration properties of B cells derived from the experiment
are such that the duration and size of the B cell-FDC contacts are sufficient to
allow for a high LT$\alpha_1\beta_2$ stimulus of a single FDC by the interaction with many B cells.
The stimulus can last for several hours which seems to be reasonable according to the
experimental data \cite{beyer:2006a,Ewijk:1980,Mackay:1998,Wang:2001}.
Thus, we conclude, that the positive
feedback loop \cite{Ansel:2000} (which is included in the simulation)
is not absolutely required to explain initial FDC formation.
\subsection{Phenomenological lymphangiogenesis model}
Lymphangiogenesis is included in the model in order to generate a LE distribution that matches the
experimental situation 
\cite{Belz:1998,Drinker:1933,Ewijk:1980,Ohtani:1986,Ohtani:1991,Belisle:1990,Pellas:1990,Belz:1995,Azzali:2000,Azzali:2002,Azzali:2003}.
It is assumed that the LE is not preformed which would determine 
shape and position of PLF a priori.
A preformed LE gap would immediately raise the question what generates this specific LE pattern.
Thus, in the simulation the generation of FDC is anti-correlated with the LE dynamics: 
LE is degraded when FDC are formed in the vicinity. It is assumed that the anti-correlation
between FDC and LE is local, i.e.~the LE-degradation process is not governed by
a long ranged diffusion of molecular messengers. 
In a similar manner LE is forming at places where FDC have transformed back
into FRC.
Angiogenesis and lymphangiogenesis models are still under development
\cite{Chaplain:2000,Chaplain:2004,Merks:2006,Namy:2004}.
Therefore, the present tissue simulation takes into account vessel formation and degradation
in the described phenomenological sense without considering the underlying mechanisms.
\subsection{Internalization of CXCR5}
The B cells have to leave the PLF
before they can exit from the SLT. 
LE will dynamically enclose the follicles. Thus, B cells frequently reach the
surface of the follicle and randomly move until a lymphatic vessel has been found for exit. Alternatively,
they might return to the follicle. It is assumed that
this search for exit points is regulated by B cell
chemotaxis. As the chemotactic response of leukocytes can be modulated by receptor internalization
(reviewed in \cite{Neel:2005}), in the simulations
this regulation is mediated by internalization of the CXCL13 receptor CXCR5.
\subsubsection{Regulation of chemotaxis}

Studies on neutrophils demonstrated that desensitization of chemoattractant receptors occurs
predominantly for high concentrations of the chemoattractant \cite{Tomhave:1994}.
One mode of desensitization is the internalization of the chemokine receptor.
However, in line with the dependence on high concentrations, it has been shown that
internalization is not absolutely required
for chemotactic responses \cite{Arai:1997,Bardi:2001,Moser:2004,Neel:2005}.
For example CCL21 does not
induce the internalization of CCR7 in contrast to CCL19 which causes the downregulation of this chemokine receptor
\cite{Bardi:2001}.
It is a general observation that different ligands of the same receptor cause different internalization levels
\cite{Neel:2005}.

Of note, some experiments fail to detect CXCL13 responses
of naive B cells freshly isolated from tonsils although the B cells where equipped with high levels
of the corresponding receptor CXCR5 \cite{Roy:2002}. Thus, the presence of chemokine receptors
CXCR5 is necessary but not sufficient to cause chemotaxis of naive B cells. 
Most likely the suppression of the function of the CXCR5 chemokine receptor
is mediated by a desensitization mechanism either by internalization
or cross-desensitization (see below).

The importance of the receptor state can be demonstrated by the response of B cells towards CCL21.
The surface CCR7 level on freshly isolated B cells is undetectable \cite{Casamayor-Palleja:2002}. The protein can
be identified in the cytoplasm and is brought to the surface in chemokine-free culture.
This suggests receptor internalization by natural exposure to CCL21 or CCL19 in the tissue.
Freshly isolated T cells also show little attraction by CCL19 \cite{Yoshida:1998}
suggesting an internalization mechanism for T cells as well. However, the CCR7 levels have not been investigated in detail.
Internalization may explain why despite the presence of CCL21 in the T zone, and lymphocyte expression
of the associated receptor CCR7, no chemotaxis
has been observed in this area \cite{Wei:2003}.

The presented data supports the notion that the chemotactic response of lymphocytes in SLT can be modified by
receptor internalization. Therefore cells may be unresponsive to chemokines despite proper receptor expression
and {\em in vitro} responses.

Note, that within the model framework regulation of chemotaxis on the basis of other mechanisms
like transcriptional regulation is not incorporated.

\subsubsection{Complex responses to multiple chemokines}
In several systems cross-talks between different chemokines/chemokine receptors have been observed
\cite{Moser:2004}.
An example is the cross-desensitization that has been reported for CXCL12 which blocks the CCL19
response of lymphocytes while
CCL19 does not block CXCL12 \cite{Kim:1998}. As expected cross-desensitization is found when 
different chemokines use the same receptor -- as shown for CCL19 and CCL21 \cite{Willimann:1998}.

Not in all cases the effect of multiple chemoattractant is pairwise blocking of the responses.
It is also possible
that the cell computes a vector sum of the incoming signals to determine an average direction of multiple
chemoattractants \cite{Foxman:1999}. Only the direction of cell migration is influenced
by the integrated response, while the speed remains unchanged. This reaction is altered
if multiple signals are given in some sequential order. Then
the direction is dominated by the newest chemoattractant even for lower concentrations and/or gradients \cite{Foxman:1999}.
An explanation for this behavior would be the desensitization of receptors for instance by internalization (see above and
\cite{Bardi:2001}). The desensitization state represents some memory for the chemoattractants a leukocyte is encountering.

Instead of desensitization a hierarchy of chemoattractants might exist. Indeed,
two signaling pathways for two different chemoattractant receptors have been identified
\cite{Heit:2002}. The hierarchy levels
are named 'target' and 'intermediary' chemoattractants. When only concurrent intermediary signals are provided
the cell responds by computing the vector sum otherwise the target chemoattractants are preferred.

Which type of response results from multiple signals may rely on the signaling cascade, i.e.~which receptors
share or use a concurring cascade to induce directed cell migration. The data available on the chemokines
CCL19, CCL21, CXCL13, and S1P, which are important for the PLF system,
are not sufficiently conclusive to allow the determination
of a signaling hierarchy. Activated naive B cells upregulate there
CCR7-levels by a factor of two to three and consequently enhance their response
to CCL19/CCL21. This balances their CXCL13 response such that they migrate to the border between the PLF and T zone
\cite{Muller:2003,Cyster:2005,Roy:2002,Casamayor-Palleja:2002,Breitfeld:2000,Reif:2002,Okada:2005}.
When either CXCR5 is overexpressed or CCR7 is lacking B cells fail to relocate to the follicle border upon antigen-stimulation
and remain in the follicle \cite{Reif:2002}. Similar if CCR7 is overexpressed by genetic manipulations the B cells
locate at the border without antigen and are moving farther into the T zone.
Analogously, activated T cells upregulate the response to CXCL13 relocating to the follicle border as well
\cite{Cyster:2005,Breitfeld:2000,Ansel:1999,Fillatreau:2003}.

Overall this suggests that the relevant chemokines in the PLF system have the
same hierarchy level and the chemotactic response of lymphocytes is a vector sum of concurrent chemokine signals.
Therefore this mode of response for CCL19, CCL21, CXCL13 and S1P has been
chosen in the model:
The speed of lymphocytes is set to a constant value derived from two-photon imaging experiments
\cite{Wei:2003,Okada:2005,Miller:2002,Miller:2003}.
The direction of the chemotactic response is the weighted average direction of the chemokine gradients.
The weight is provided by the gradient of the bound chemokine at both ends of the cell
assuming that the cell senses the difference of bound molecules across its diameter.
\section{Results}
\bfig
 \centering
 \includegraphics{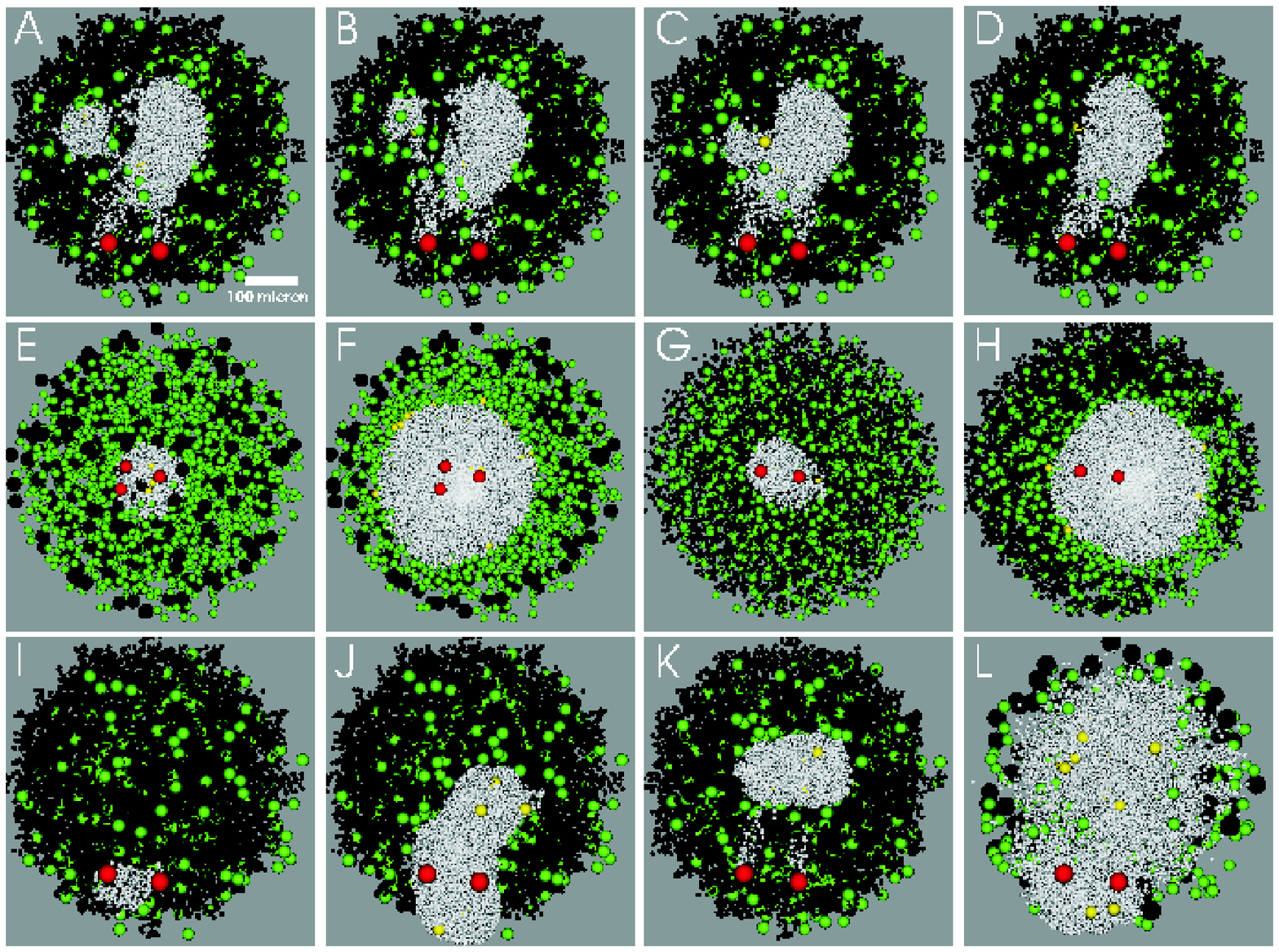}
 \caption{\label{fig-morph}
          Three dimensional slice projections from simulations (rendered with POVRay, http://www.povray.org).
          Objects are shown as colored spheres (yellow: FDC, green: FRC, white: naive B cells, red: HEV, dark grey: ELV).
            The internalization process destabilizes
            the follicle shape, even when the FDC network is of fixed size and ELV are distributed
            in the T zone (A--D, interval between images is 1 h).
            Also the dynamics of the efferent lymphatic destabilizes the follicle shape (E: 5 h, F: 122 h, G: 5 h, H: 57 h).
            Although the morphology of the
            follicle is correct with the LE surrounding a spherical follicle the situation does not equilibrate.
            This is independent of the density of the LE (E,F: low density; G,H: high density).
            Note that at high LE density the distinct spheres representing the exit spots merge into a seemingly homogeneous
            background in the pictures.
            Introducing negative regulation with free FDC dynamics the follicle forms around the
            HEV (I: 10 h, J: 60 h). If however, like in the shown case, B cells chemotax towards CCL21 the follicle
            may elongate towards the center of the stromal network (J).
            Simulating B cell homeostasis with negative regulation of FDC generation
            and a preformed FDC network with high density LE
            the follicle rapidly adopts its final shape and keeps its position and does not
            relocate to the HEV (K: 5 days).
            When S1P is acting in the dynamic PLF formation a disturbed follicular border occurs (L: 1 day).
            Also the PLF appears large due to the lower density of B cells while the number of B cells is not
            significantly increased (not shown). 
          }
\efig
Simulations are performed on the basis of the model 
as introduced in the previous section.
The minimal set of assumptions entering the model are 
\begin{itemize}
\item fixed number of non-migrating stromal cells,
\item constant entrance of B cells via small number of HEV,
\item B cells use the LE for egress from the SLT,
\item generation of FDC by B cell LT$\alpha_1\beta_2$ signaling,
\item replacement of FRC by generated FDC,
\item removal of FDC (replaced by FRC) in case of lack of LT$\alpha_1\beta_2$ stimulation,
\item secretion of CXCL13 by FDC,
\item secretion of CCL21 by FRC,
\item secretion of S1P by LE,
\item chemotactic activity of B cells to CXCL13, CCL21, and S1P,
\item B cells leaving the SLT via LE when S1P$_1$ levels are sufficient to dominate CXCL13 response.
\end{itemize}
These assumptions alone lead to reasonable cell dynamics. Restricting
the LE to a small area generates a PLF that is stable in
size, shape, and position (see \cite{beyer:2006a})
using physiological values for the parameters
(Table \ref{tab-par}).
However, the position of the follicle is in contradiction
to microanatomical data
\cite{Belz:1998,Drinker:1933,Ewijk:1980,Ohtani:1986,Ohtani:1991,Belisle:1990,
Pellas:1990,Belz:1995,Azzali:2000,Azzali:2002,Azzali:2003}.
Introducing the correct relative position of PLF and LE
with dynamic LE, dramatically changes the follicle stability.
Thus, the stability of follicle shape and size in an
equilibrium of cell flow crucially depends on further mechanisms.

In the following possible mechanisms are discussed that guarantee
the stability of the follicle. The most intuitive hypothesis,
is related to internalization dynamics of chemotaxis receptors.
It is shown that this hypothesis leads to unphysiological results
and is therefore unlikely to be the relevant factor of
physiological follicle formation and maintenance.
Realistic results could only be achieved by the assumption
of a so far unknown mechanism of negative regulation of
FDC generation. This mechanism is a prediction of the model.

\subsection{Including internalization dynamics}
The internalization of CXCR5 (modeled by Eq.~\ref{eq-reacsys}
and \cite{beyer:2006a})
destabilizes the shape of the follicle defined by the B cells
(Fig.~\ref{fig-morph}, A--D).
In a first step, the dynamics of the FDC network was switched off 
in order to isolate the effect of B cell internalization dynamics.

The instability is caused by uptake of CXCL13
when the CXCR5-CXCL13 complex is internalized: 
The chemotactically responding B cells act as a sink
for CXCL13. This generates steep gradients in the 
chemokine distribution which guide the B cell movement.
Interestingly, these local gradients, induced by the
responding lymphocyte population itself,
can become so strong that they reverse the chemotaxis
gradient. Then cell movement is not strictly directed
towards the chemokine source anymore.

In principle, a large diffusion 
constant could counter-balance this tendency. However,
the reversal of chemokine gradients is most pronounced 
at larger distances of the chemokine source.
An unphysiologically large
diffusion constant would be required to counter-act 
the local dynamics in the chemokine concentration.

Internalization can induce quasi-periodical
alterations of the chemokine concentrations.
Assume a population of B cells with full CXCR5 expression 
and high concentration of CXCL13. The receptor
will be internalized on all cells using up the CXCL13 
to some extent. The lowered CXCL13 concentration and
CXCR5 levels promote B cells to enter a random walk 
migration modus and spread. The low uptake of CXCL13 by the B cells
with internalized CXCR5 permits the CXCL13 concentration to rise again. 
The combination of CXCL13 uptake and switch of cell migration between
chemotaxis and random migration 
leads to a local synchronization of the receptor dynamics 
of these cells. The spreading of the B cells destabilizes the shape
of the PLF and becomes significant when local populations of 
B cells synchronously spread out.

The instability of the B cell aggregate leads 
to high B cell densities outside the FDC network.
If the FDC network dynamics is switched on again
this culminates in the generation of new FDC at the border of the PLF.
The area covered by the B cells becomes more extended than expected
from a densely packed set of B cells in the FDC network 
(see also \cite{beyer:2006a}).
In other words, the volume of the FDC network becomes bigger than 
the volume required by the number of B cells. 
Thus, the whole PLF becomes unstable in shape and follows the shifted
'center of mass' of the FDC network
as a result of the CXCL13 concentration peaks. 

The positive feedback loop (CXCL13 stimulating higher LT$\alpha_1\beta_2$ levels
in B cells, inducing more CXCL13 secreting FDC \cite{Ansel:2000})
enhances this effect because B cells at low density 
can already induce FDC when their LT$\alpha_1\beta_2$ levels
are high due to the CXCL13 stimulus. This is most pronounced 
when the peak LT$\alpha_1\beta_2$ levels of B cells are delayed compared to the
peak CXCL13 stimulus because then the cells have a high probability 
to enter a FDC free area with enhanced surface LT$\alpha_1\beta_2$ levels.

Note that in a previous study no instability was induced by
internalization dynamics \cite{beyer:2006a}. This is
related to S1P chemotaxis towards the LE and to a wrong
position of the LE in this study: The follicle formed around the
LE position. 

In conclusion, internalization dynamics of CXCR5 
induce unstable PLF. As such instabilities are not observed
in nature, the present model predicts that 
receptor internalization is not
a dominant process in PLF formation and maintenance. Even though
it can not be excluded that internalization is active
and only attenuated by further interactions, it is
unlikely that we loose essential features of PLF formation
when neglecting internalization dynamics in the following.
However, the robustness of the results with respect to
internalization dynamics will be considered.

\subsection{Including lymphatic endothelium dynamics}

\bfig
 \centering
 \includegraphics[width=7cm,angle=-90]{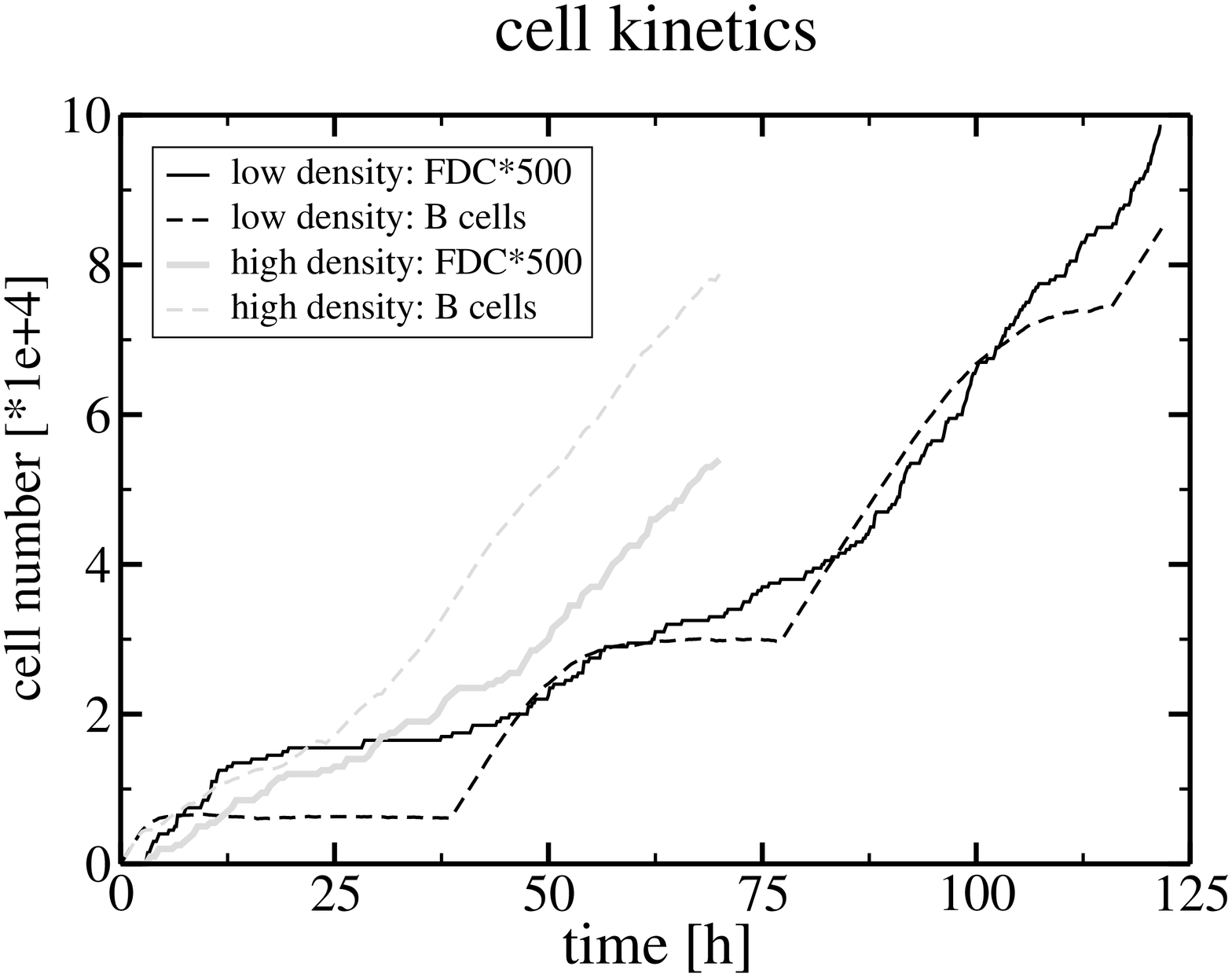}
 \caption{\label{fig-kin}
            The dynamics of the efferent lymphatic destabilizes  the follicle shape.
            The limiting factor for the follicle growth is mainly the influx of B cells as
            shown by the number of cells at each time point. In the case of low density LE there
            exist metastable states where the B cell number is almost constant while the number of
            FDC is growing until it reaches the LE and pushes it away which then leads to
            increased B cell numbers. The number of FDC has been multiplied by
            a factor 500 for reasons of clear visibility. } 
\efig

LE dynamics might be considered to stabilize follicle
shape and size by locating the exit points for B cells.
However, LE dynamics turn out to destabilizes the follicle 
system independently of the internalization dynamics 
(Fig.~\ref{fig-morph}, E--H and Fig.~\ref{fig-kin}).
The reason for this instability is as follows: 
B cells that approach the LE -- either by random
motion or S1P-directed migration \refsec{sec-s1p} 
-- can stimulate FRC to become FDC if they occur at sufficient
density such that FDC are generated right next to the exit spots.
As FDC and LE are anti-correlated, 
i.e.~the vicinity of LE is free of FDC, the generation of FDC
leads to the degradation of the LE. The exit spots 
for B cells are pushed further away. 
The B cells need to approach the LE and 'follow' the
distant exit spots. Due to the increased distance of the exit spots from the
follicle center the B cells take longer to reach 
the spots thus increasing their transit time. Consequently
the follicle is enlarged and the density of B cells 
close to the exit spots again reaches concentrations
to induce new FDC. Thus a self-perpetuating process drives 
the border between FDC and LE leading to a constantly
growing PLF size.

Even though the model for the genesis of lymph endothelium is rather
simplistic, it nevertheless covers a reasonable idea
of the dynamics of LE. The destabilizing effect of this
basic model points towards an additional process
that prevents the PLF from constant growth
by inhibition of the generation of FDC at the right
follicle size.
\subsection{Negative regulation of FDC generation}
A negative regulation mechanism that inhibits the 
generation of FDC can be realized in two distinct ways.
First, the FDC induction is prevented when some factor 
is missing, i.e.~FDC precursor activation gets lost.
That implies that cells located
in the T zone but not in the follicle secrete a 
substance that keeps the FDC progenitors in a state
in which they are 
able to become FDC.
The production and sensitivity for the signal has to be
such that only many cells are able to give rise
to a sufficient signal concentration.
The fact that
the T zone can be absent while separated PLF
are observable in certain knockout experiments 
\cite{Fu:1999} makes this scenario rather
unlikely.

The second option is that the generation of FDC leads 
to the secretion of a substance X that inhibits
their own generation counter-balancing the effect of LT$\alpha_1\beta_2$.
Note that the negative regulation shall not affect 
existing FDC and not induce their dedifferentiation to FRC.
Also the signal X has to be diffusible. 
A local inhibition mechanism might prevent
the generation of FDC but does it independently of the follicle
size. A signal sensitive to the follicle size needs to be accumulating 
and dispersing the information in the vicinity of the PLF.

The simulations show that negative regulations of FDC induction, indeed,
leads to stable follicles. Both aforementioned scenarios have similar
effects (Fig.~\ref{fig-morph}, I and J).
The PLF forms close to the HEV as expected from
previous work where a small area containing LE had been 
positioned in a relatively large distance to the HEV \cite{beyer:2006a}.
In this configuration the PLF were forming around 
the LE and the HEV. In the present simulation the 
formation around the LE is no longer possible, because the 
homogeneous distribution of LE does not provide an 
aggregation center for the B cells to reach sufficient high density for induction
of FDC via a large LT$\alpha_1\beta_2$-stimulus. The lack of a B cell aggregation center
in the present work is rather 
independent on S1P acting chemotactically on B cells or not. The broad distribution
of S1P-producing LE can only provide shallow gradients not leading to B cell aggregation.
The chemotactic activity of B cells in response to CCL21 can however lead
to elongated follicles when the a CCL21 concentration peak is nearby the
PLF center (Fig.~\ref{fig-morph}, J).

The presence of one of these negative regulation mechanism has impact on the
morphology of the PLF now exhibiting two zones. One B cell rich zone
with FDC and without vessels, and a shell of B cells with 
low or vanishing FDC density but with LE.
Within this shell B cells are exiting from the PLF 
and SLT. 

An additional stability test can be performed using a preformed FDC
network that is placed in a certain distance from the HEV. The simulation
of this configurations shows that the preformed FDC network rapidly
extents to its final size and most importantly remains at its location
(Fig.~\ref{fig-morph}, K). Thus, in contrast to simulations in \cite{beyer:2006a},
the B cells passing through the LE
and the FRC network are not able to induce FDC before approaching in the PLF.
\subsection{Why S1P chemotaxis is not likely to occur in primary lymphoid follicles}\label{sec-s1p}
An important result of this study is that the chemotaxis 
towards S1P is no longer required and even
more that S1P chemotaxis might not be active in the PLF system.
In contrast to the previous study \cite{beyer:2006a} 
the S1P chemotaxis is not required for B cells
to find an exit spot because the LE is distributed 
homogeneously around the follicle. This makes it easy
for B cells to find the LE by random migration within a rather short time.
Thus, the role of a chemical signal has been replaced by a specific
functional morphology.

The effect of S1P in the PLF system is mainly 
a disturbed border of the lymphoid follicle
(Fig.~\ref{fig-morph}, L). This results
from a rather low concentration of CXCL13 at 
the boundary of the follicle which allows S1P to influence
the direction of motion of B cells at the follicle border. 
This is only slightly affected by the
internalization dynamics of CXCR5 because the cells at 
the border are mainly sensitive for CXCL13 due to
low CXCL13 concentration. 
The morphology caused by S1P suggests
that B cells in the follicle are not chemotactically responding to S1P.
\section{Discussion}
An agent-based lattice-free simulation tool for
the formation and maintenance of PLF was presented.
The generation of FDC is
induced by B cells providing sufficient LT$\alpha_1\beta_2$ signals 
to stromal progenitor cells. B cells are sensitive
to chemokine CXCL13 via their receptor CXCR5.
CXCL13 is secreted by FDC. The positive feedback loop
that enhance B cell LT$\alpha_1\beta_2$ levels by CXCL13 stimulation \cite{Ansel:2000}
is not required to understand PLF formation.
Optionally,
the desensitization of the CXCR5 receptor via 
ligand-induced receptor internalization is considered.
A simple model for the microanatomy
of the lymphatic vessels in the vicinity of lymphoid 
follicles is used assuming an anti-correlation
between LE and FDC.  

In contrast to previous simulations \cite{beyer:2006a},
which did not consider CXCR5 internalization
dynamics or LE anti-correlation with FDC,
PLF are found to be unstable when at least one
of these additional features is included.
In the following this is explored in more detail,
and a mechanism is proposed to elevate the
problem of follicle instability. 
This model prediction is discussed in the context
of related experiments.

{\it Internalization of chemokine receptors does not drive PLF formation}\\
The instability generated by the internalization dynamics 
of CXCR5 on B cells in the simulation suggests that this process might not
be relevant in the PLF system. 
The internalization of CXCR5 has been introduced in the simulation 
to desensitize chemotactic attraction of B cells to the center of 
the FDC network and, thus, to make them approach
the follicular border either by random migration or chemotactic 
migration in response to other chemokines like CCL21, CCL19,
and/or S1P, and, ultimately, to allow B cells to leave the follicle.
A couple of points
argue against this process being relevant to B cells leaving
the follicle.

In order to compensate for shape fluctuations induced by the CXCR5
internalization dynamics and achieve a spherical PLF the production
of CXCL13 could in principle be raised. The resulting CXCL13 concentration
has to be high enough to ensure that CXCL13 removal by CXCR5 internalization
is not strong enough to invert the CXCL13 gradient.
However, the high CXCL13 concentration
would result in high CXCL13 concentration everywhere
in the PLF leading to rapidly CXCR5 desensitized B cells even at the border
of the PLF. Consequently
CXCL13 would not be able to hold B cells for a long time
and B cells would leave the follicle after 15--30 minutes \cite{Neel:2005}.
This short PLF transit time is in contradiction to the
experiments observing transit of at least 3--5 hours
\cite{Pellas:1990,Pellas:1990b,Young:1999,Srikusalanukul:2002}. 

Another argument against the internalization dynamics is that
it cannot be stable in regulating the size of follicle.
Assume that by random fluctuation more B cells arrive at the 
follicle. Internalization reduces the CXCL13 levels
such that cells are less desensitized and remain longer in the 
follicle. As a consequence the follicle is growing in B cell numbers.
The generation of FDC (with and without the positive
feedback loop \cite{Ansel:2000}) is by far slower
than cell migration such that the B cells number remains at a
higher ratio compared to the CXCL13 sources and letting 
the follicle grow further.
Similar arguments apply to the case when a fluctuation leads to lower B cell numbers.
When less B cells are in the follicle, then CXCL13 is 
dominating and B cells leave the follicle fast by a more rapid
and frequent CXCR5 internalization. The CXCL13 concentration 
in the shrinking follicle is rising further until the slower dynamics
of the FDC reduces the CXCL13 amount. However, due to their faster 
dynamics the number of B cells decreases faster. Basically, the
slow dynamics of the FDC compared to the migration and internalization 
dynamics of B cells is responsible for this instability.

Note, that it cannot be ruled out that a another 
modification of the CXCR5 receptor levels,
for example on the transcriptional level, leads to a desensitization
of B cells for CXCL13 thus enabling them to leave the PLF.
Also it cannot be excluded that mechanisms exist that compensate for the
internalization induced instability of the PLF.
This may be a desired situation considering that the internalization dynamics
can induce cycling of naive B cells between the center of the PLF
and its border. The cycling is due to the strong desensitization in the
center of the PLF where high CXCL13 concentrations exist. B cells can then approach the
border by random migration where the lower CXCL13 concentrations permit
a resensitization for CXCL13 chemotaxis. These B cells migrate back to the center
of the PLF.

{\it Follicular B cells are not responsive to CCL19, CCL21, and S1P}\\
The morphological data suggest that B cells in the PLF are not 
chemotactical responsive for S1P. A superposition of the two
chemoattractants S1P and CXCL13 generates a blurry PLF border 
which is not observed experimentally. This is in line with the recent
observation that S1P is not a chemotactic factor {\em in vivo} for 
T cells in the LN \cite{Wei:2005}.
The same blurry PLF border would be induced if
follicular B cells responded to the T zone chemokines CCL19 and CCL21.
Therefore, the simulation suggests
that chemotaxis to CCL19, CCL21 and S1P 
is suppressed in follicular B cells, which seems to be the case
{\em in vivo} \cite{Cyster:2005,Willimann:1998,Okada:2002,Rangel-Moreno:2005} 
despite the positive response {\em in vitro}
\cite{Kim:1998,Yoshida:1998}.
This may be caused by internalization of CCR7 
at the border of the PLF where the concentrations of CCL19
and CCL21 are still quite high \cite{Okada:2005,Luther:2002} which
is supported by the missing CCR7 surface levels of freshly 
isolated B cells \cite{Casamayor-Palleja:2002}.

{\it A CXCL13-LT$\alpha_1\beta_2$ positive feedback loop initiates PLF formation and gradual FDC differentiation}\\
The function of the positive feedback loop \cite{Ansel:2000} 
(LT$\alpha_1\beta_2$ levels are increased by CXCL13, and LT$\alpha_1\beta_2$ induce
novel FDC which are the source of CXCL13)
was analyzed in the simulations. There are two possible
functions: First, a randomly formed B cell aggregate may induce 
a minimum level of CXCL13 that starts the feedback loop 
leading to increased LT$\alpha_1\beta_2$ and finally to increased CXCL13.
The CXCL13 levels reached in this way are sufficiently strong 
to induce chemotaxis in B cells and to dominate other chemotactic 
responses (e.g.~to CCL21). This initiates formation of follicles.
However, {\em in vitro} data show that very high CXCL13 levels
(10 nM) are required to efficiently
induce LT$\alpha_1\beta_2$ on B cells \cite{Luther:2002}. These concentrations 
are already optimal for chemotaxis
\cite{Legler:1998,Gunn:1998}, which sets in question the
former functional interpretation of the feedback loop.
If one, nevertheless, assumes that
the positive feedback loop is required to induce 
initial CXCL13 expression, then the initiation of the first FDC
requires either another helper cell type or high LT$\alpha_1\beta_2$ levels 
on some B cells. The latter might be a result of stimulation with
antigen \cite{Ansel:2000}. Then, either
the cellular helper source needs to be identified or the 
dependence on B cell activation has to be confirmed.

An alternative interpretation is that the LT$\alpha_1\beta_2$ signal resulting
solely from the aggregation of B cells may be 
sufficient to induce CXCL13-producing FDC but not other FDC markers
like CD21 or CD35 which require the high LT$\alpha_1\beta_2$ levels 
achieved by the feedback loop \cite{Ansel:2000}.
The gradual maturation of FDC 
observed in other studies  
\cite{Bofill:2000,Kasajima-akatsuka:2006}
supports this notion. 

Moreover the notion of gradual maturation favors the strict 
interpretation of FDC as excited state of stromal cells
and speaks against the differentiation of an unknown progenitor. 
This may also explain why in certain experiments
FDC are rapidly lost when the LT$\alpha_1\beta_2$ stimulus is blocked with antibodies
\cite{Mackay:1998,Wang:2001,Gommerman:2002,Huber:2005}. 
Thus the loss of the LT$\alpha_1\beta_2$ signal may not lead
to the loss of FDC by apoptosis but a 'decay' 
of exited stromal cell back to their native state.

{\it Lymphatic vessels}\\
The microanatomical location of the LE has been modeled by 
an exclusion principle: Whenever a FDC is newly appearing the
lymphatic vessels in the vicinity are removed. 
Presumably, remodeling is organized 
by the exchange of chemicals inducing behavioral changes 
of the LE cell. This is not considered explicitly but
summarized in a time delay of LE remodeling 
of several hours. 
Analogously the dedifferentiation of a FDC into a FRC
is followed by the creation of LE. 

The regulation of the LE is not precisely understood.
In the following several hypothetic pathways that may
regulate the LE dynamics and their corresponding experimental
evidence are discussed.
The LE dynamics could be related to the ECM. The reticular fibers are sparse in the follicle
\cite{Ohtani:1991,Ohtsuka:1992,Castanos-velez:1995,Ushiki:1995,Katakai:2004,Satoh:1997,Liakka:1992,Maatta:2004}
and their degradation in the PLF could cause the disintegration of lymphatic vessels.
A candidate signal to be involved 
is the expression of matrix metalloproteinases
by FDC \cite{Braun:2004}.
The generation of ECM components by FRC provides the ground for 
LE cells to extent the network in this area. 
The required vascular endothelial growth factors VEGF-C and/or VEGF-D 
can be provided from cells like macrophages 
(which can transdifferentiate into LE cells themselves) 
\cite{Kerjaschki:2005} or dendritic cells \cite{Hamrah:2003}.
In this view, the gradual destruction of the reticular network in 
developing follicles in sheep \cite{Halleraker:1994} would be
accompanied by a gradual loss of LE
in this area, which, in turn, promotes PLF growth.

There exists indirect evidence how LE remodeling might be regulated in the PLF.
Angiogenesis and lymphangiogenesis 
during airway infections is dependent on B cells \cite{Aurora:2005}.
A candidate for the mediator signal is vascular endothelial growth factor (VEGF)-D \cite{Aurora:2005}.
Although the cellular source is not clear VEGF-D could
directly couple to B cells as suggested by the stimulation of VEGF
by c-Myc$^+$ B cells in lymphoma \cite{Ruddell:2003}.
A VEGF source candidate in SLT are DC in the T zone \cite{Hamrah:2003}.
The co-localization of DC in the T zone correlates with the
presence of LE. Thus it may turn out that DC are important for
the structure of SLT. The lack of LE in the PLF may then
be coupled to the absence of DC and not to a LE-destructing process
initiated by the FDC. This hypothesis is already covered
by the present simulations
outcome of the simulation because in the real tissue
as soon as FDC form, the DC are
replaced by B cells. Thus VEGF will no longer be 
present and consequently lymphatic
vessels are not maintained.
An experiment using a system with VEGF$^{-/-}$ DC may
elucidate the role of DC in SLT lymphangiogenesis.

Assuming that DC are responsible for the location of LE,
the ring of B cells found in certain LT$\alpha_1\beta_2$ and/or TNF-$\alpha$-deficiencies
\cite{Fu:1999,Muller:2003,Tumanov:2003} should be devoid of LE.
However, the absence of LE in the B cell ring does not exclude the possibility
that high B cells densities lead to the reduction of LE. Thus, the anti-correlation
of FDC and LE assumed in the simulation can be re-interpreted
to be indirect in the way that FDC support the aggregation of B cells leading
to reduced LE in the PLF.

TNF-$\alpha$ might be not only involved in the generation of FDC \cite{Fu:1999,Muller:2003,Tumanov:2003}
but directly in the LE remodeling process.
Both B and T cells can provide TNF-$\alpha$ \cite{Fu:1999,Tumanov:2003}.
TNF-$\alpha$ can upregulate VEGF-C
(\cite{Frater-Schroder:1987} and references in \cite{Hamrah:2003}).
The anti-correlation between FDC and vessels may
be related to the TNF-$\alpha$ levels in the PLF that are raised
by TNF-$\alpha$-producing FDC \cite{Corcione:1997}. It is known that TNF-$\alpha$ can
also have anti-angiogenic effects at high concentrations \cite{Patterson:1996}.
TNF-$\alpha$ in the PLF may be present at sufficiently high
levels to induce anti-angiogenic effects .

The presented model assumes that the lymphatic network formation
in SLT is coupled to the dynamics of PLF. In contrast one
could assume that the LE is formed before PLF formation
and determines the position and shape of the developing PLF.
A direct test for the existence of the LE dynamics is
to provide anti-angiogenic factors to LT$\alpha_1\beta_2^{-/-}$ or similar
knockouts prior to reconstitution by wildtype lymphocytes
or bone marrow. The PLF that form should be much
smaller than usual and contain LE. Alternatively, the
inverse experiment could be done by blocking LT$\alpha_1\beta_2$ in wildtype
mice or reconstitute them with bone marrow from LT$\alpha_1\beta_2^{-/-}$.
The disruption of the PLF should be followed by the presence
of vessels in these areas. When using anti-angiogenic factors during the experiment,
the gaps in the vessel network due to the PLF should be preserved when the PLF disappear.

{\it A novel mechanism of negative regulation of FDC generation}\\
The instabilities created by the LE dynamics require a
mechanism that negatively regulates the induction of FDC.
Even though negative regulation might be realized
in a number of different mechanisms,
from the simulation viewpoint, only the source and the
propagation of the signal initiating negative regulation are relevant.
Either a signal has to be reduced or to be
produced when FDC are generated. In both cases the 
signal overcomes a threshold at the border of the PLF.
The signal is required to be diffusive 
in order to act on the length scale of a PLF
and can be shown to require a rather fast 
decay in order to establish a time-independent (stable) 
follicle size (Appendix \ref{sec-app2}).

Within the framework of the simulation an inhibiting 
signal is suggested that is produced when FDC are generated.
Note that this inhibiting signal may be secreted 
either by B cells located in the PLF or by FDC.
If the B cells are the sources of the hypothetic signal,
they would constitutively produce the substance such that small
B cell aggregates induce FDC while larger ones inhibit
FDC induction. Alternatively, B cells sense the presence
of FDC
in the PLF (e.g~via CXCL13) and produce the inhibitor 
upon stimulation in concurrence to LT$\alpha_1\beta_2$. 

The constitutive production of the inhibiting signal
by B cells implies
that the PLF would always grow from low B cell
numbers to its final size and large aggregates 
that formed due to some other process would prohibit follicle
formation. 
Considering the PLF formation in mice 
lacking T cells \cite{Fu:1999} this seems an unlikely
situation because the B cell dense tissue clearly shows 
PLF formation. Also in normal MALT
and LN organogenesis an intermediate
ring of B cells around the developing T zone appears 
\cite{Mebius:1997,Nishikawa:2003,Finke:2005,Fu:1999,Muller:2003,Yoshida:1999,Hashi:2001,Kim:2000,Cupedo:2004}.
The ring formation 
suggests that high B cell numbers 
do not suppress FDC generation. Thus most likely
either only the B cells in the follicle get stimulated
to produce a negatively regulating 
factor or the FDC themselves are the source.

{\it Primary lymphoid follicles exhibit two zones}\\
The regulation of the follicle size under the 
influence of the dynamics of efferent lymphatic vessels
implies that the B cell follicle consists of two
zones. One zone contains B cells and the FDC network 
the other one B cells and efferent lymphatic vessels. From the
corrosion casts this organization cannot be determined 
\cite{Azzali:2000,Azzali:2002,Azzali:2003} but studies with
FDC marker seem to support this view
\cite{Muller:2003,Ansel:2000,Kasajima-akatsuka:2006,Cupedo:2004,Balogh:2001,Fu:1997,Fu:1998,Gommerman:2002}.
However, in the framework of the present simulation
the FDC network is defined by the CXCL13 distribution.
Thus, the state of seemingly absent FDC in the outer zone of the PLF
is not resolved in the model. The results may also
indicate that the state of 'FDC-ness' 
decreases radially towards the border of the follicle.

{\it A theory of PLF formation and maintenance}\\
The simulation of PLF formation and maintenance gives rise
to the following vantage point:
FDC are a differentiated state of stromal precursor cells.
Differentiation is promoted by LT$\alpha_1\beta_2$, which is provided
by B cells.
The initiation of PLF formation is related to
small aggregates of B cells and does not rely on the
a positive feedback loop \cite{Ansel:2000} between
CXCL13 and LT$\alpha_1\beta_2$.
B cells in a follicle have to desensitize their chemotactic
responsiveness to T zone derived chemokines (CCL19, CCL21)
in order to establish a realistic flow of B cells through
the follicle and to avoid follicle instabilities. The model predicts
that a factor negatively regulating FDC generation determines 
the size of a PLF. This factor stabilizes PLF shape and size. 
It is most likely 
an inhibitory diffusing signal secreted from either
FDC or B cells residing in the PLF.

A drawback of the presented model is that other desensitization mechanisms
for the chemotactic response of B cells are not considered due to a lack
of corresponding data. Other regulatory pathways like the transcription
of the CXCR5 receptor may modify the PLF formation theory drawn in this
study.
The lymphangiogenesis model used here is very simple and only phenomenological.
A detailed modeling of the lymphangiogenesis during PLF formation
is left for future research, especially the use of a real vessel
structure instead of spherical representatives. The dynamics of the
adjacent T zone may influence PLF formation, although
preliminary results seem not to change the conclusion drawn
from the presented results. In addition to a more detailed
lymphangiogenesis the dynamics
of the HEV during PLF formation might be considered.
\section{Method}
\label{sec-method}
The PLF system is simulated using an agent-based off-lattice model
\cite{Odell:1981,Palsson:2000,Dallon:2004,Drasdo:1995,Drasdo:2004,Honda:2004,Schaller:2005}.
To achieve a lattice-free description a regular triangulation has been chosen
\cite{Schaller:2004,Schaller:2005,beyer:2005,beyer:2006a}.
The regular triangulation, which is a generalization of the Delaunay triangulation,
is used to provide the neighborhood topology for the cells that allows for
a continuous representation of cell positions and sizes.
The simulations of cells is represented in a 3-level model. 
The first level is the internal state of the
cells representing the dynamics of the phenotype of the cell. 
The second level models the contact interaction
between cells including mechanical interactions with 
the environment and exchange of signals by membrane
bound molecules like LT$\alpha_1\beta_2$. The third level incorporates 
long range interactions via diffusive substances
which is used to describe the chemokine distribution in the PLF.
The corresponding equations are briefly summarized in this section. 
More details of the model can be found in \cite{beyer:2006a}.

\subsection{Internal cell dynamics}
The phenotype of a cell is described by a set of 
internal cell variables. These include internal times
to indicate when which type of event may happen. 
An example is the persistence time of cells during chemotactic
motion \cite{Wei:2003,Okada:2005,Miller:2002,Miller:2003,Gunzer:2004,mehe:2005}.
When the persistence time has past, the cell can reorient its migration direction
to the local chemokine field \cite{Albrecht:1998,Ehrengruber:1996}.
Another variable is the differentiation of 
a FRC to a FDC where two times are important: The time
the LT$\alpha_1\beta_2$ stimulus is provided and the time 
the differentiation is completed (see below).

A set of variables describes the mechanics of the cell, 
e.g.~velocity, cell polarization, and cell volume.
These variable couple directly to the next level 
of description, the contact interaction of cells, and have no
influence on other internal states directly but can 
influence the variables of neighbor cells.
Additionally there are parameters like the 
elasticity constant mentioned below that can be cell-type specific.
Further details of the model and an extensive 
set of parameters have been published \cite{beyer:2006a}
and can be found in the supplementary material.

Upon contact, two cells can exchange signals via their
contact surface. In a simple approach with unpolarized
ligand/receptor distributions the signal 
strength is proportional to the contact area
provided by the Voronoi tessellation
\cite{Meineke:2001,Schaller:2005,beyer:2006a}.

\subsubsection{Internal dynamics of FRC and FDC}
Cell differentiation is described as phenotype change of a cell.
The realization of cell differentiation is derived 
from the change of cell behavior and internal states
upon internal or external stimuli. 
Within the model a FRC differentiates to a FDC 
when the signal threshold for LT$\alpha_1\beta_2$ has been exceeded for
a given time $T_{\rm FRC\leftarrow FDC}$. The signal 
is determined by summing up all LT$\alpha_1\beta_2$ contributions
from neighbor cells, i.e.~surface density of LT$\alpha_1\beta_2$ times 
contact area. The differentiation is then instantly
performed changing the internal cell states 
of a FRC into that of a FDC. In a similar manner FDC
differentiate back to FDC after the LT$\alpha_1\beta_2$ signal is 
below the threshold for a given time $T_{\rm FDC \leftarrow FRC}$.
To reduce the amount of parameters and considering the lack of experimental data
$T_{\rm FRC \leftarrow FDC} = T_{\rm FDC \leftarrow FRC}$ is assumed.
Also the thresholds for LT$\alpha_1\beta_2$ are identical
constants for FRC to FDC and FDC to FRC differentiation.

The positive feedback loop \cite{Ansel:2000} involving B cells is realized by a linear interpolation of
surface LT$\alpha_1\beta_2$ level between a constitutive low level and CXCL13-induced high LT$\alpha_1\beta_2$ level. The induction by CXCL13
is chosen to be proportional to the fraction of bound CXCR5 receptor implying an immediate response.
As an additional variant the LT$\alpha_1\beta_2$ level can been chosen proportional to the amount of internalized CXCR5
to model a delayed response. Additional delays which may result by the transcription of LT$\alpha_1\beta_2$ have not been considered.
This seems to be realistic considering that B cells reside in the PLF for
a few hours \cite{Pellas:1990,Pellas:1990b,Young:1999,Srikusalanukul:2002} which permits high LT$\alpha_1\beta_2$ levels only
when the cells are about to leave the follicle, if at all.
\subsection{Equations of motion}
The contact interaction of cells is primarily 
the mechanical interaction between cells. It is described by
Newtonian equations of motion in the overdamped 
approximation.
In  this approximation acceleration of
cells and consequently conservation of moment can be ignored. The equations incorporate elastic
cell responses upon cell contact, forces generated by actively migrating cells, and friction
forces between cells as well as cells and extracellular matrix.

The Newtonian equations of motion read
\be\label{eq-newton}
 \begin{aligned}
  m_i\mathbf {\ddot x_i} &= \mathbf F^{\rm act}_i(\phi_i)
                  + \sum_{j \in \mathcal N_i}\left[
                                             \mathbf F^{\rm act}_{ij}(\phi_i) - \mathbf F^{\rm act}_{ji}(\phi_j)
                                            + \mathbf F^{\rm pass}_{ij}\right] + \mathbf F^{\rm drag}_i \approx 0.
 \end{aligned}
\ee
The forces at position $x_i$ of the cell $i$ depend on the internal state $\phi$ of the cells for
active forces $\mathbf F^{\rm act}$ and on the cell's position as well as on the position of all
neighbor cells $j \in \mathcal N_i $ for passive forces $\mathbf F^{\rm pass}$.
These forces are counter-balanced by the velocity-dependent drag forces $\mathbf F^{\rm drag}$ resulting in a ODE system of first order for the cell positions.

The passive forces are the elastic described by the JKR-model \cite{Johnson:1971}. It depends on the virtual cell
overlap $h_{ij} = r_i + r_j - d_{ij}$ where $d_{ij}$ is the center distance of the cells and
$r_i$ and $r_j$ are the cell radii.
\be\label{eq-JKR}
 \begin{aligned}
  \mathbf F^{\rm pass}_{ij}\left(\mathbf x_i,\mathbf x_j\right)
                     &= \left\{E_{ij}^* \sqrt{r_{ij}^*}\,h_{ij}^{3/2} - \sqrt{6\pi \sigma_{ij}E_{ij}^*{r_{ij}^*}^{3/2} h_{ij}^{3/2}}\right\}
                        \mathbf {\hat e}_{ij}\\
  \frac{1}{E_{ij}^*} &= \frac{3}{4}\left[\frac{1-\nu_i^2}{E_i} + \frac{1-\nu_j^2}{E_j}\right]\\
  \frac{1}{r_{ij}^*} &= \frac{1}{r_i} + \frac{1}{r_j}
  \end{aligned}
\ee
with cell elasticity constants $E_i$ and $E_j$, Poisson numbers $\nu_i$ and $\nu_j$.

The force acting on cell $i$ by exerting active forces on a neighbor cell $j$ reads
\be
  \mathbf F^{\rm act}_{ij}(\phi_i) = a_{ij} p^*_i \,\mathrm{sign}([\mathbf x^*_{ij} - \mathbf x^*_i]\cdot\mathbf o_i)
                             \,\frac{\mathbf x^*_{ij} - \mathbf x^*_i}{\|\mathbf x^*_{ij} - \mathbf x^*_i\|}
\ee
with the cell orientation $\mathbf o_{ij}$, cell surface contact point $\mathbf x^*_{ij}$, constriction ring
center $\mathbf x^*_i$, interaction area
$a_{ij}$ and the pressure $p^*_i$ exerted by cell i.
Additionally a constant active force $-\mathbf F^{\rm act}_i$ is directly exerted by the cell on the ECM by the cell
driving the cell in the direction of the cell axis $\mathbf o_i$.

The drag force is linear in the velocity $\mathbf v$ given by
\be\label{eq-drag}
 \begin{aligned} 
  \mathbf F^{\rm drag}_i &= 
    -\eta_{\rm med}r_i\left(1 - \frac{A}{A^{\rm tot}_i}\right) \mathbf v_i\\
    &\quad+ \sum_{j\in \mathcal N_i} \left(\eta_i r_i+\eta_j r_j\right)\frac{a_{ij}}{A^{\rm tot}_i}
    \left[\mathbf v_{ij}  -  \mathbf {\hat e}_{ij}
    \left(\mathbf {\hat e}_{ij}\cdot\mathbf v_{ij}\right) \right]
 \end{aligned}
\ee
with specific coefficients $\eta_i$, $\eta_j$, and $\eta_{\rm medium}$.
$A^{\rm tot}_i$ is the total surface of a cell and $A = \sum_j a_{ij}$ is the surface in contact with other cells.
\subsection{Reaction-diffusion system of chemokines}
The chemotaxis of cells is described by coupling the direction of the active force of a cell to the
local chemokine gradient. According to the observation that leukocytes tend to have a persistence time
between subsequent orientation changes \cite{Wei:2003,Okada:2005,Miller:2002,Miller:2003}
the gradient is sensed by the simulated cells periodically.
The concentration of the chemokines CXCL13, and CCL21 are computed solving the time dependent diffusion equation.
CCL19 is not explicitly calculated as it has a similar distribution like CCL21 \cite{Cyster:2005}, acts on the same receptor
\cite{Yoshida:1998}, and has a far lower concentration that CCL21 \cite{Luther:2002}.
Note, that in inhomogeneous media like complex tissue a reaction-diffusion approach can become invalid
and refinements might be necessary \cite{Schnell:2004}.

The receptor dynamics of the free ($R$) and internalized receptor ($R^*$) are described by the following
equations:
\be\label{eq-reacsys}
 \begin{aligned}
  \dot R         \quad&=\quad - k_{\rm on}R\,c + k_{\rm off}(R_{\rm tot} - R - R^*) + k_{\rm r} R^*\\
  \dot R^*       \quad&=\quad k_{\rm i}(R_{\rm tot} - R - R^*) - k_{\rm r}R^*\\
  \dot c         \quad&=\quad - k_{\rm on}R\,c + k_{\rm off}(R_{\rm tot} - R - R^*) - \kappa c + Q.
 \end{aligned}
\ee
$R_{\rm tot}$ represents the constant number of receptors in any state of a cell. $k_{\rm on}$ and
$k_{\rm off}$ are the rate constants for the ligand receptor binding. The variable $c$ describes the
concentration of the ligand which decays with the constant $\kappa$ and is produced by the
source $Q$. $k_{\rm i}$ and $k_{\rm r}$ are the rate constants for internalization of the receptor-ligand
complexes and the recycling of receptors, respectively.

In the model the LE is assumed to attract lymphocytes via the chemoattractant S1P. For simplicity and based
on the notion that the dynamics of the LE is coupled to a slow process -- the FDC dynamics -- and by itself is a slow
process it is calculated using the Poisson equation. Thus the time dependence of the S1P concentration is assumed to play
no role and is omitted.
Similarly the internalization dynamics of the S1P receptor is kept simple without explicit internalization dynamics
as presented in \refeq{eq-reacsys}.
The reexpression of S1P$_1$ is made independent of the S1P levels and increases linearly until the levels permits the cell
to enter LE after about 3 hr \cite{Pellas:1990,Pellas:1990b,Young:1999,Srikusalanukul:2002,Lo:2005}.
\subsection{Initial conditions}
For most of the simulations a stromal background consisting of CCL21 producing FRC with a centrally localized preformed FDC
network has been used. The area containing the FRC contains a constant density of exit spots which resemble the LE
where cells can leave the tissue.
A fixed number of HEV is located acentrically outside the preformed FDC network. The HEV are represented as spheres where
cells enter the tissue. The majority of the simulations is performed with an influx of B cells only.

\begin{table}
 \centering
 {\small
 \begin{tabular}{lrll} \hline
   {\bf parameter} & {\bf value} & & {\bf remarks/references} \\ \hline
   B/T cell diameter                   & 9 \MICRON~    & & \cite{Ewijk:1980,Thompson:1984,Bornens:1989,Haston:1984} \\
   $E_i$               & 1 kPa & & \cite{Bausch:1998,Bausch:1999,Forgacs:1998,Chu:2005} \\
   $\nu_i$               & 0.4   & & \cite{Maniotis:1997,Hategan:2003} \\
   $\sigma_{ij}$          & 0--0.3 ${\rm nN}\,\mu{\rm m}^{-1}$ & & \cite{Moy:1999,Verdier:2003} \\
   $F^{\rm act}_i$     & 50--200 nN & & \cite{Balaban:2001,Burton:1999,Schwarz:2002b} \\
   $p^*_i$ & 0.4 ${\rm nN}\,\mu{\rm m}^2$ & & \cite{Galbraith:1999} \\
   $T_{\rm p}$                  & 120--180 s & & \cite{Miller:2002,Wei:2003} \\
   $\eta_i,\,\eta_{\rm med}$             & 1500 nN$\,\mu{\rm m}^{-1}\,s$ & & \cite{Miller:2002,Miller:2003,Wei:2003,Okada:2005,Bausch:1998,Bausch:1999,Lauffenburger:1996}\\
    
   \hline
   \LT~threshold               & 1                   & & arbitrary units \\
   low density \LT         & $0.025\, \mu{\rm m}^{-2}$ & & unknown\\
   high density \LT        & $0.5\, \mu{\rm m}^{-2}$    & & unknown\\
   $T_{\rm FRC \rightarrow FDC}$ & 3 h & & \cite{Ewijk:1980,Mackay:1998,Mebius:1997} \\ 
   
   \hline
   
   $D$                              & 10--100 $\mu{\rm m}^2\,s^{-1}$ & & \cite{Randolph:2005} \\
   size of diffusion grid                          & 1200\MICRON~& & \\
   grid resolution                    & 35 \MICRON~ & & \\
   max.~cell displacement $\Delta x$ & 0.9\MICRON~  & &  \\
   min.~time resolution $\Delta t$   & 10 s         & & \\
   B  : T cell ratio                & 0.4 : 0.6    & & \cite{Young:1999,Sacca:1998}\\
   influx of cells (B + T)              & 0.1--2 ${\rm cells}\,s^{-1}$ & & \cite{Young:1999,Sacca:1998} \\
   size of simulation area          & 600 \MICRON~ & & \\
   number of FRC                        & 2500         & & \\
   
   \hline
   $k_{\rm i}$  & $5 \cdot 10^{-5} \ldots 3\cdot10^{-2} \,{\rm s^{-1}}$ & &\cite{Neel:2005} \\
   $k_{\rm r}$  & $1 \cdot 10^{-4} \ldots 7\cdot10^{-3} \,{\rm s^{-1}}$ & &\cite{Neel:2005} \\
   $K_{\rm d}$  & $0.2\ldots 5\,{\rm nM}$                               & &\cite{Willimann:1998,Yoshida:1998,Lin:2004,Pelletier:2000,Slimani:2003} \\
   $k_{\rm on}$ & $2.5\cdot10^5 \ldots 10^8 \,{\rm M^{-1}s^{-1}} $      & &\cite{Pelletier:2000} \\
   $k_{\rm off}$ & $10^{-4} \ldots 1 \,{\rm s^{-1}}  $                  & &(from $K_{\rm d}$ and $k_{\rm on})$ \\
   $Q$          & $2.5\cdot 10^1\ldots 10^4\,{\rm s^{-1}}$              & &\cite{Vissers:2001,Hu:2004} \\ 
   $R_{\rm tot}$                 & $10^4$--$10^5$ & & \cite{Willimann:1998,Okada:2002} \\
   
   \hline
 \end{tabular}
 }
 \vspace{0.3cm}
 \caption{\label{tab-par} The parameters used in the simulation of the PLF.
          The references given support the used value.
          If no reference or comment, the parameter is a systemic model 
          parameter chosen to get sufficient accuracy in the simulation.}
\end{table}

\begin{appendix}

\section{Estimating the life time of a signal for the negative regulation of FDC generation}\label{sec-app2}
Solving the stationary diffusion equation with a decay term and a homogeneous
spherical source $f$ of size $L$ (spherical coordinates with radius $\rho$)
\be
  0 = D\left(\frac{\partial^2 c}{\partial \rho^2} + 2\rho\frac{\partial c}{\partial \rho} \right) - \kappa c -f\, \Theta(L-\rho)
\ee
yields the solution
\be
 \begin{aligned}
 c &= \frac{f}{\kappa}\Theta(L-\rho)
     + \frac{f}{\kappa}\Theta(\rho-L)\left[ \frac{L}{\rho} \cosh{\frac{\rho-L}{l}}
       + \frac{l}{\rho} \sinh{\frac{\rho-L}{l}}\right]\\
   &\quad -\frac{f}{\kappa}\left(\frac{L}{l}+1\right)\frac{e^{-L/l}}{\rho/l}\sinh{\frac{\rho}{l}}
 \end{aligned}
\ee
with $l=\sqrt{D/\kappa}$ the typical length scale of the system. The negative regulation should set on around the
coordinate $L$ where the source of the negative signal vanishes. The highest values at this border will be reached when
$L \gg l$. However, the PLF growth will stop as soon as this threshold is reached such that the typical radius of the PLF
will be in the order of a few $l$. For typical parameters of the diffusion coefficient of $10\ldots100\,\mu{\rm m}^2s^{-1}$
the value for $\kappa$ should be in the range of $10^{-3}\ldots10^{-2}\,\rm s$, i.e.~the typical lifetime of the molecule
mediating the negative regulation is in the order of minutes to tens of minutes.
The decay term may be just the uptake of the signal by the cells it is acting on or its enzymatic inactivation.

\end{appendix}

\section*{Acknowledgement}
The ALTANA AG supports the Frankfurt Institute for Advanced Studies (FIAS).


\begin{thebibliography}{100}
\expandafter\ifx\csname url\endcsname\relax
  \def\url#1{{\tt #1}}\fi
\expandafter\ifx\csname urlprefix\endcsname\relax\def\urlprefix{URL }\fi
\providecommand{\eprint}[2][]{\url{#2}}

\bibitem{Maclennan:1994}
I.~C. MacLennan (1994).
\newblock {G}erminal centers.
\newblock {\em Annu.~Rev.~Immunol.\/}, {\bf 12}:117--39.

\bibitem{Kosco-vilbois:2003}
M.~H. Kosco-Vilbois (2003).
\newblock {A}re follicular dendritic cells really good for nothing?
\newblock {\em Nat.~Rev.~Immunol.\/}, {\bf 3}(9):764--9.

\bibitem{Nicander:1991}
L.~Nicander, M.~Halleraker, and T.~Landsverk (1991).
\newblock {O}ntogeny of reticular cells in the ileal {P}eyer's patch of sheep
  and goats.
\newblock {\em Am.~J.~Anat.\/}, {\bf 191}(3):237--49.

\bibitem{Griebel:1996}
P.~J. Griebel and W.~R. Hein (1996).
\newblock {E}xpanding the role of {P}eyer's patches in {B}-cell ontogeny.
\newblock {\em Immunol.~Today\/}, {\bf 17}(1):30--9.

\bibitem{Makala:2002}
L.~H. Makala, N.~Suzuki, and H.~Nagasawa (2002-2003).
\newblock {P}eyer's patches: organized lymphoid structures for the induction of
  mucosal immune responses in the intestine.
\newblock {\em Pathobiology\/}, {\bf 70}(2):55--68.

\bibitem{Mebius:1997}
R.~E. Mebius, P.~Rennert, and I.~L. Weissman (1997).
\newblock Developing lymph nodes collect {CD}4+{CD}3- {LT}$\beta$+ cells that
  can differentiate to {APC}, {NK} cells, and follicular cells but not {T} or
  {B} cells.
\newblock {\em Immunity\/}, {\bf 7}(4):493--504.

\bibitem{Nishikawa:2003}
S.~Nishikawa, K.~Honda, P.~Vieira, and H.~Yoshida (2003).
\newblock {O}rganogenesis of peripheral lymphoid organs.
\newblock {\em Immunol.~Rev.\/}, {\bf 195}:72--80.

\bibitem{Finke:2005}
D.~Finke (2005).
\newblock {F}ate and function of lymphoid tissue inducer cells.
\newblock {\em Curr.~Opin.~Immunol.\/}, {\bf 17}(2):144--50.

\bibitem{Fu:1999}
Y.~X. Fu and D.~D. Chaplin (1999).
\newblock {D}evelopment and maturation of secondary lymphoid tissues.
\newblock {\em Annu.~Rev.~Immunol.\/}, {\bf 17}:399--433.

\bibitem{Muller:2003}
G.~Muller, U.~E. Hopken, and M.~Lipp (2003).
\newblock {T}he impact of {CCR}7 and {CXCR}5 on lymphoid organ development and
  systemic immunity.
\newblock {\em Immunol.~Rev.\/}, {\bf 195}:117--35.

\bibitem{Tumanov:2003}
A.~V. Tumanov, S.~I. Grivennikov, A.~N. Shakhov, S.~A. Rybtsov, E.~P. Koroleva,
  J.~Takeda, S.~A. Nedospasov, and D.~V. Kuprash (2003).
\newblock {D}issecting the role of lymphotoxin in lymphoid organs by
  conditional targeting.
\newblock {\em Immunol Rev\/}, {\bf 195}:106--16.

\bibitem{Cyster:2005}
J.~G. Cyster (2005).
\newblock {C}hemokines, sphingosine-1-phosphate, and cell migration in
  secondary lymphoid organs.
\newblock {\em Annu.~Rev.~Immunol.\/}, {\bf 23}:127--59.

\bibitem{Mackay:1998}
F.~Mackay and J.~L. Browning (1998).
\newblock {T}urning off follicular dendritic cells.
\newblock {\em Nature\/}, {\bf 395}(6697):26--7.

\bibitem{Ansel:2000}
K.~M. Ansel, V.~N. Ngo, P.~L. Hyman, S.~A. Luther, R.~Forster, J.~D. Sedgwick,
  J.~L. Browning, M.~Lipp, and J.~G. Cyster (2000).
\newblock {A} chemokine-driven positive feedback loop organizes lymphoid
  follicles.
\newblock {\em Nature\/}, {\bf 406}(6793):309--14.

\bibitem{Garside:1998}
P.~Garside, E.~Ingulli, R.~R. Merica, J.~G. Johnson, R.~J. Noelle, and M.~K.
  Jenkins (1998).
\newblock {V}isualization of specific {B} and {T} lymphocyte interactions in
  the lymph node.
\newblock {\em Science\/}, {\bf 281}(5373):96--9.

\bibitem{Bhalla:1981}
D.~K. Bhalla, T.~Murakami, and R.~L. Owen (1981).
\newblock {M}icrocirculation of intestinal lymphoid follicles in rat {P}eyer's
  patches.
\newblock {\em Gastroenterology\/}, {\bf 81}(3):481--91.

\bibitem{Ohtani:2003}
O.~Ohtani, Y.~Ohtani, C.~J. Carati, and B.~J. Gannon (2003).
\newblock {F}luid and cellular pathways of rat lymph nodes in relation to
  lymphatic labyrinths and {A}quaporin-1 expression.
\newblock {\em Arch.~Histol.~Cytol.\/}, {\bf 66}(3):261--72.

\bibitem{Halleraker:1994}
M.~Halleraker, C.~M. Press, and T.~Landsverk (1994).
\newblock {D}evelopment and cell phenotypes in primary follicles of foetal
  sheep lymph nodes.
\newblock {\em Cell.~Tissue Res.\/}, {\bf 275}(1):51--62.

\bibitem{Brachtel:1996}
E.~F. Brachtel, M.~Washiyama, G.~D. Johnson, K.~Tenner-Racz, P.~Racz, and I.~C.
  MacLennan (1996).
\newblock {D}ifferences in the germinal centres of palatine tonsils and lymph
  nodes.
\newblock {\em Scand.~J.~Immunol.\/}, {\bf 43}(3):239--47.

\bibitem{Kasajima-akatsuka:2006}
N.~Kasajima-Akatsuka and K.~Maeda (2006).
\newblock {D}evelopment, maturation and subsequent activation of follicular
  dendritic cells ({FDC}): immunohistochemical observation of human fetal and
  adult lymph nodes.
\newblock {\em Histochem.~Cell.~Biol.\/}, pages 1--13.

\bibitem{Belz:1998}
G.~T. Belz (1998).
\newblock {I}ntercellular and lymphatic pathways associated with tonsils of the
  soft palate in young pigs.
\newblock {\em Anat.~Embryol.~(Berl)\/}, {\bf 197}(4):331--40.

\bibitem{Kumar:2006}
P.~Kumar and J.~F. Timoney (2006).
\newblock {H}istology, immunohistochemistry and ultrastructure of the tonsil of
  the soft palate of the horse.
\newblock {\em Anat.~Histol.~Embryol.\/}, {\bf 35}(1):1--6.

\bibitem{beyer:2006a}
T.~Beyer and M.~Meyer-Hermann (2006).
\newblock Modeling emergent tissue organization involving high-speed migrating
  cells in a flow equilibrium.
\newblock {\em submitted to Phys.~Rev.~E\/}.

\bibitem{Schaller:2004}
G.~Schaller and M.~Meyer-Hermann (2004).
\newblock {K}inetic and dynamic {D}elaunay tetrahedralizations in three
  dimensions.
\newblock {\em Comput.~Phys.~Commun.\/}, {\bf 162}:9--23.

\bibitem{Schaller:2005}
G.~Schaller and M.~Meyer-Hermann (2005).
\newblock Multicellular tumor spheroid in an off-lattice voronoi-delaunay cell
  model.
\newblock {\em Phys.~Rev.~E.\/}, {\bf 71}(5 Pt 1):051910.

\bibitem{beyer:2005}
T.~Beyer, G.~Schaller, A.~Deutsch, and M.~Meyer-Hermann (2005).
\newblock {P}arallel dynamic and kinetic regular triangulation in three
  dimensions.
\newblock {\em Comput.~Phys.~Commun.\/}, {\bf 172}(2):86--108.

\bibitem{Wei:2005}
S.~H. Wei, H.~Rosen, M.~P. Matheu, M.~G. Sanna, S.~K. Wang, E.~Jo, C.~H. Wong,
  I.~Parker, and M.~D. Cahalan (2005).
\newblock {S}phingosine 1-phosphate type 1 receptor agonism inhibits
  transendothelial migration of medullary {T} cells to lymphatic sinuses.
\newblock {\em Nat.~Immunol.\/}, {\bf 6}(12):1228--35.

\bibitem{Drinker:1933}
C.~K. Drinker, G.~B. Wislocki, and M.~E. Field (1933).
\newblock {T}he structure of the sinuses in the lymph nodes.
\newblock {\em Anat.~Rec.\/}, {\bf 56}(3):261--273.

\bibitem{Ewijk:1980}
W.~van Ewijk and T.~H. van~der Kwast (1980).
\newblock {M}igration of {B} lymphocytes in lymphoid organs of lethally
  irradiated, thymocyte-reconstituted mice.
\newblock {\em Cell.~Tissue Res.\/}, {\bf 212}(3):497--508.

\bibitem{Ohtani:1986}
O.~Ohtani, A.~Ohtsuka, and R.~L. Owen (1986).
\newblock {T}hree-dimensional organization of the lymphatics in the rabbit
  appendix. {A} scanning electron and light microscopic study.
\newblock {\em Gastroenterology\/}, {\bf 91}(4):947--55.

\bibitem{Ohtani:1991}
O.~Ohtani, A.~Kikuta, A.~Ohtsuka, and T.~Murakami (1991).
\newblock {O}rganization of the reticular network of rabbit {P}eyer's patches.
\newblock {\em Anat.~Rec.\/}, {\bf 229}(2):251--8.

\bibitem{Belisle:1990}
C.~Belisle and G.~Sainte-Marie (1990).
\newblock {B}lood vascular network of the rat lymph node: tridimensional
  studies by light and scanning electron microscopy.
\newblock {\em Am.~J.~Anat.\/}, {\bf 189}(2):111--26.

\bibitem{Pellas:1990}
T.~C. Pellas and L.~Weiss (1990).
\newblock {D}eep splenic lymphatic vessels in the mouse: a route of splenic
  exit for recirculating lymphocytes.
\newblock {\em Am.~J.~Anat.\/}, {\bf 187}(4):347--54.

\bibitem{Belz:1995}
G.~T. Belz and T.~J. Heath (1995).
\newblock {I}ntercellular and lymphatic pathways of the canine palatine
  tonsils.
\newblock {\em J.~Anat.\/}, {\bf 187 (Pt1)}:93--105.

\bibitem{Azzali:2000}
G.~Azzali and M.~L. Arcari (2000).
\newblock {U}ltrastructural and three dimensional aspects of the lymphatic
  vessels of the absorbing peripheral lymphatic apparatus in {P}eyer's patches
  of the rabbit.
\newblock {\em Anat.~Rec.\/}, {\bf 258}(1):71--9.

\bibitem{Azzali:2002}
G.~Azzali, M.~Vitale, and M.~L. Arcari (2002).
\newblock {U}ltrastructure of absorbing peripheral lymphatic vessel ({ALPA}) in
  guinea pig {P}eyer's patches.
\newblock {\em Microvasc.~Res.\/}, {\bf 64}(2):289--301.

\bibitem{Azzali:2003}
G.~Azzali (2003).
\newblock {S}tructure, lymphatic vascularization and lymphocyte migration in
  mucosa-associated lymphoid tissue.
\newblock {\em Immunol.~Rev.\/}, {\bf 195}:178--89.

\bibitem{Ekino:1979}
S.~Ekino, K.~Matsuno, and M.~Kotani (1979).
\newblock {D}istribution and role of lymph vessels of the bursa {F}abricii.
\newblock {\em Lymphology\/}, {\bf 12}(4):247--52.

\bibitem{Irjala:2001}
H.~Irjala, E.~L. Johansson, R.~Grenman, K.~Alanen, M.~Salmi, and S.~Jalkanen
  (2001).
\newblock {M}annose receptor is a novel ligand for {L}-selectin and mediates
  lymphocyte binding to lymphatic endothelium.
\newblock {\em J.~Exp.~Med.\/}, {\bf 194}(8):1033--42.

\bibitem{Matloubian:2004}
M.~Matloubian, C.~G. Lo, G.~Cinamon, M.~J. Lesneski, Y.~Xu, V.~Brinkmann, M.~L.
  Allende, R.~L. Proia, and J.~G. Cyster (2004).
\newblock {L}ymphocyte egress from thymus and peripheral lymphoid organs is
  dependent on {S}1{P} receptor 1.
\newblock {\em Nature\/}, {\bf 427}(6972):355--60.

\bibitem{Sallusto:2004}
F.~Sallusto and C.~R. Mackay (2004).
\newblock {C}hemoattractants and their receptors in homeostasis and
  inflammation.
\newblock {\em Curr.~Opin.~Immunol.\/}, {\bf 16}(6):724--31.

\bibitem{Lo:2005}
C.~G. Lo, Y.~Xu, R.~L. Proia, and J.~G. Cyster (2005).
\newblock {C}yclical modulation of sphingosine-1-phosphate receptor 1 surface
  expression during lymphocyte recirculation and relationship to lymphoid organ
  transit.
\newblock {\em J.~Exp.~Med.\/}, {\bf 201}(2):291--301.

\bibitem{Pellas:1990b}
T.~C. Pellas and L.~Weiss (1990).
\newblock {M}igration pathways of recirculating murine {B} cells and {CD}4+ and
  {CD}8+ {T} lymphocytes.
\newblock {\em Am.~J.~Anat.\/}, {\bf 187}(4):355--73.

\bibitem{Young:1999}
A.~J. Young (1999).
\newblock {T}he physiology of lymphocyte migration through the single lymph
  node in vivo.
\newblock {\em Semin.~Immunol.\/}, {\bf 11}(2):73--83.

\bibitem{Srikusalanukul:2002}
W.~Srikusalanukul, F.~De~Bruyne, and P.~McCullagh (2002).
\newblock {A}n application of linear output error modelling for studying
  lymphocyte migration in peripheral lymphoid tissues.
\newblock {\em Australas.~Phys.~Eng.~Sci.~Med.\/}, {\bf 25}(3):132--8.

\bibitem{Crivellato:1997}
E.~Crivellato and F.~Mallardi (1997).
\newblock {S}tromal cell organisation in the mouse lymph node. {A} light and
  electron microscopic investigation using the zinc iodide-osmium technique.
\newblock {\em J.~Anat.\/}, {\bf 190 ( Pt 1)}:85--92.

\bibitem{Kaldjian:2001}
E.~P. Kaldjian, J.~E. Gretz, A.~O. Anderson, Y.~Shi, and S.~Shaw (2001).
\newblock {S}patial and molecular organization of lymph node {T} cell cortex: a
  labyrinthine cavity bounded by an epithelium-like monolayer of fibroblastic
  reticular cells anchored to basement membrane-like extracellular matrix.
\newblock {\em Int.~Immunol.\/}, {\bf 13}(10):1243--53.

\bibitem{Nolte:2003}
M.~A. Nolte, J.~A. Belien, I.~Schadee-Eestermans, W.~Jansen, W.~W. Unger,
  N.~van Rooijen, G.~Kraal, and R.~E. Mebius (2003).
\newblock {A} conduit system distributes chemokines and small blood-borne
  molecules through the splenic white pulp.
\newblock {\em J.~Exp.~Med.\/}, {\bf 198}(3):505--12.

\bibitem{Balogh:2004}
P.~Balogh, G.~Horvath, and A.~K. Szakal (2004).
\newblock {I}mmunoarchitecture of distinct reticular fibroblastic domains in
  the white pulp of mouse spleen.
\newblock {\em J.~Histochem.~Cytochem.\/}, {\bf 52}(10):1287--98.

\bibitem{Nierop:2002}
K.~van Nierop and C.~de~Groot (2002).
\newblock {H}uman follicular dendritic cells: function, origin and development.
\newblock {\em Semin.~Immunol.\/}, {\bf 14}(4):251--7.

\bibitem{Bofill:2000}
M.~Bofill, A.~N. Akbar, and P.~L. Amlot (2000).
\newblock {F}ollicular dendritic cells share a membrane-bound protein with
  fibroblasts.
\newblock {\em J.~Pathol.\/}, {\bf 191}(2):217--26.

\bibitem{Szakal:1985}
A.~K. Szakal, R.~L. Gieringer, M.~H. Kosco, and J.~G. Tew (1985).
\newblock {I}solated follicular dendritic cells: cytochemical antigen
  localization, {N}omarski, {SEM}, and {TEM} morphology.
\newblock {\em J.~Immunol.\/}, {\bf 134}(3):1349--59.

\bibitem{Clark:1992}
E.~A. Clark, K.~H. Grabstein, and G.~L. Shu (1992).
\newblock {C}ultured human follicular dendritic cells. {G}rowth characteristics
  and interactions with {B} lymphocytes.
\newblock {\em J.~Immunol.\/}, {\bf 148}(11):3327--35.

\bibitem{Chang:2003}
K.~C. Chang, X.~Huang, L.~J. Medeiros, and D.~Jones (2003).
\newblock {G}erminal centre-like versus undifferentiated stromal
  immunophenotypes in follicular lymphoma.
\newblock {\em J.~Pathol.\/}, {\bf 201}(3):404--12.

\bibitem{Tsunoda:1997}
R.~Tsunoda, A.~Bosseloir, K.~Onozaki, E.~Heinen, K.~Miyake, H.~Okamura,
  K.~Suzuki, T.~Fujita, L.~J. Simar, and N.~Sugai (1997).
\newblock {H}uman follicular dendritic cells in vitro and follicular
  dendritic-cell-like cells.
\newblock {\em Cell Tissue Res.\/}, {\bf 288}(2):381--9.

\bibitem{Cyster:2000}
J.~G. Cyster, K.~M. Ansel, K.~Reif, E.~H. Ekland, P.~L. Hyman, H.~L. Tang,
  S.~A. Luther, and V.~N. Ngo (2000).
\newblock {F}ollicular stromal cells and lymphocyte homing to follicles.
\newblock {\em Immunol.~Rev.\/}, {\bf 176}:181--93.

\bibitem{Kapasi:1998}
Z.~F. Kapasi, D.~Qin, W.~G. Kerr, M.~H. Kosco-Vilbois, L.~D. Shultz, J.~G. Tew,
  and A.~K. Szakal (1998).
\newblock {F}ollicular dendritic cell ({FDC}) precursors in primary lymphoid
  tissues.
\newblock {\em J.~Immunol.\/}, {\bf 160}(3):1078--84.

\bibitem{Kuprash:2005}
D.~V. Kuprash, A.~V. Tumanov, D.~J. Liepinsh, E.~P. Koroleva, M.~S. Drutskaya,
  A.~A. Kruglov, A.~N. Shakhov, E.~Southon, W.~J. Murphy, L.~Tessarollo, S.~I.
  Grivennikov, and S.~A. Nedospasov (2005).
\newblock {N}ovel tumor necrosis factor-knockout mice that lack {P}eyer's
  patches.
\newblock {\em Eur.~J.~Immunol.\/}, {\bf 35}(5):1592--600.

\bibitem{Pasparakis:2000}
M.~Pasparakis, S.~Kousteni, J.~Peschon, and G.~Kollias (2000).
\newblock {T}umor necrosis factor and the p55{TNF} receptor are required for
  optimal development of the marginal sinus and for migration of follicular
  dendritic cell precursors into splenic follicles.
\newblock {\em Cell.~Immunol.\/}, {\bf 201}(1):33--41.

\bibitem{Lee:2005}
S.~Lee, K.~Choi, H.~Ahn, K.~Song, J.~Choe, and I.~Lee (2005).
\newblock {T}uj1 (class {III} beta-tubulin) expression suggests dynamic
  redistribution of follicular dendritic cells in lymphoid tissue.
\newblock {\em Eur.~J.~Cell Biol.\/}, {\bf 84}(2-3):453--9.

\bibitem{Crivellato:2004}
E.~Crivellato, A.~Vacca, and D.~Ribatti (2004).
\newblock {S}etting the stage: an anatomist's view of the immune system.
\newblock {\em Trends Immunol.\/}, {\bf 25}(4):210--7.

\bibitem{Wang:2001}
Y.~Wang, J.~Wang, Y.~Sun, Q.~Wu, and Y.~X. Fu (2001).
\newblock {C}omplementary effects of {TNF} and lymphotoxin on the formation of
  germinal center and follicular dendritic cells.
\newblock {\em J.~Immunol.\/}, {\bf 166}(1):330--7.

\bibitem{Chaplain:2000}
M.~A. Chaplain (2000).
\newblock {M}athematical modelling of angiogenesis.
\newblock {\em J.~Neurooncol.\/}, {\bf 50}(1-2):37--51.

\bibitem{Chaplain:2004}
M.~Chaplain and A.~Anderson (2004).
\newblock {M}athematical modelling of tumour-induced angiogenesis: network
  growth and structure.
\newblock {\em Cancer Treat.~Res.\/}, {\bf 117}:51--75.

\bibitem{Merks:2006}
R.~M. Merks, S.~V. Brodsky, M.~S. Goligorksy, S.~A. Newman, and J.~A. Glazier
  (2006).
\newblock Cell elongation is key to in silico replication of in vitro
  vasculogenesis and subsequent remodeling.
\newblock {\em Dev.~Biol.\/}, {\bf 289}(1):44--54.

\bibitem{Namy:2004}
P.~Namy, J.~Ohayon, and P.~Tracqui (2004).
\newblock {C}ritical conditions for pattern formation and in vitro
  tubulogenesis driven by cellular traction fields.
\newblock {\em J.~Theor.~Biol.\/}, {\bf 227}(1):103--20.

\bibitem{Neel:2005}
N.~F. Neel, E.~Schutyser, J.~Sai, G.~H. Fan, and A.~Richmond (2005).
\newblock {C}hemokine receptor internalization and intracellular trafficking.
\newblock {\em Cytokine Growth Factor Rev.\/}, {\bf 16}(6):637--58.

\bibitem{Tomhave:1994}
E.~D. Tomhave, R.~M. Richardson, J.~R. Didsbury, L.~Menard, R.~Snyderman, and
  H.~Ali (1994).
\newblock {C}ross-desensitization of receptors for peptide chemoattractants.
  {C}haracterization of a new form of leukocyte regulation.
\newblock {\em J.~Immunol.\/}, {\bf 153}(7):3267--75.

\bibitem{Arai:1997}
H.~Arai, F.~S. Monteclaro, C.~L. Tsou, C.~Franci, and I.~F. Charo (1997).
\newblock {D}issociation of chemotaxis from agonist-induced receptor
  internalization in a lymphocyte cell line transfected with {CCR}2{B}.
  {E}vidence that directed migration does not require rapid modulation of
  signaling at the receptor level.
\newblock {\em J.~Biol.~Chem.\/}, {\bf 272}(40):25037--42.

\bibitem{Bardi:2001}
G.~Bardi, M.~Lipp, M.~Baggiolini, and P.~Loetscher (2001).
\newblock {T}he {T} cell chemokine receptor {CCR}7 is internalized on
  stimulation with {ELC}, but not with {SLC}.
\newblock {\em Eur.~J.~Immunol.\/}, {\bf 31}(11):3291--7.

\bibitem{Moser:2004}
B.~Moser, M.~Wolf, A.~Walz, and P.~Loetscher (2004).
\newblock {C}hemokines: multiple levels of leukocyte migration control.
\newblock {\em Trends Immunol.\/}, {\bf 25}(2):75--84.

\bibitem{Roy:2002}
M.~P. Roy, C.~H. Kim, and E.~C. Butcher (2002).
\newblock {C}ytokine control of memory {B} cell homing machinery.
\newblock {\em J.~Immunol.\/}, {\bf 169}(4):1676--82.

\bibitem{Casamayor-Palleja:2002}
M.~Casamayor-Palleja, P.~Mondiere, C.~Verschelde, C.~Bella, and T.~Defrance
  (2002).
\newblock {BCR} ligation reprograms {B} cells for migration to the {T} zone and
  {B}-cell follicle sequentially.
\newblock {\em Blood\/}, {\bf 99}(6):1913--21.

\bibitem{Yoshida:1998}
R.~Yoshida, M.~Nagira, M.~Kitaura, N.~Imagawa, T.~Imai, and O.~Yoshie (1998).
\newblock {S}econdary lymphoid-tissue chemokine is a functional ligand for the
  {CC} chemokine receptor {CCR}7.
\newblock {\em J.~Biol.~Chem.\/}, {\bf 273}(12):7118--22.

\bibitem{Wei:2003}
S.~H. Wei, I.~Parker, M.~J. Miller, and M.~D. Cahalan (2003).
\newblock {A} stochastic view of lymphocyte motility and trafficking within the
  lymph node.
\newblock {\em Immunol.~Rev.\/}, {\bf 195}:136--59.

\bibitem{Kim:1998}
C.~H. Kim, L.~M. Pelus, J.~R. White, E.~Applebaum, K.~Johanson, and H.~E.
  Broxmeyer (1998).
\newblock {CK}$\beta$-11/macrophage inflammatory protein-3$\beta$/{EBI}1-ligand
  chemokine is an efficacious chemoattractant for {T} and {B} cells.
\newblock {\em J.~Immunol.\/}, {\bf 160}(5):2418--24.

\bibitem{Willimann:1998}
K.~Willimann, D.~F. Legler, M.~Loetscher, R.~S. Roos, M.~B. Delgado,
  I.~Clark-Lewis, M.~Baggiolini, and B.~Moser (1998).
\newblock {T}he chemokine {SLC} is expressed in {T} cell areas of lymph nodes
  and mucosal lymphoid tissues and attracts activated {T} cells via {CCR}7.
\newblock {\em Eur.~J.~Immunol.\/}, {\bf 28}(6):2025--34.

\bibitem{Foxman:1999}
E.~F. Foxman, E.~J. Kunkel, and E.~C. Butcher (1999).
\newblock {I}ntegrating conflicting chemotactic signals. {T}he role of memory
  in leukocyte navigation.
\newblock {\em J.~Cell.~Biol.\/}, {\bf 147}(3):577--88.

\bibitem{Heit:2002}
B.~Heit, S.~Tavener, E.~Raharjo, and P.~Kubes (2002).
\newblock {A}n intracellular signaling hierarchy determines direction of
  migration in opposing chemotactic gradients.
\newblock {\em J.~Cell.~Biol.\/}, {\bf 159}(1):91--102.

\bibitem{Breitfeld:2000}
D.~Breitfeld, L.~Ohl, E.~Kremmer, J.~Ellwart, F.~Sallusto, M.~Lipp, and
  R.~Forster (2000).
\newblock {F}ollicular {B} helper {T} cells express {CXC} chemokine receptor 5,
  localize to {B} cell follicles, and support immunoglobulin production.
\newblock {\em J.~Exp.~Med.\/}, {\bf 192}(11):1545--52.

\bibitem{Reif:2002}
K.~Reif, E.~H. Ekland, L.~Ohl, H.~Nakano, M.~Lipp, R.~Forster, and J.~G. Cyster
  (2002).
\newblock {B}alanced responsiveness to chemoattractants from adjacent zones
  determines {B}-cell position.
\newblock {\em Nature\/}, {\bf 416}(6876):94--9.

\bibitem{Okada:2005}
T.~Okada, M.~J. Miller, I.~Parker, M.~F. Krummel, M.~Neighbors, S.~B. Hartley,
  A.~O'Garra, M.~D. Cahalan, and J.~G. Cyster (2005).
\newblock {A}ntigen-engaged {B} cells undergo chemotaxis toward the {T} zone
  and form motile conjugates with helper {T} cells.
\newblock {\em PLoS.~Biol.\/}, {\bf 3}(6):e150.

\bibitem{Ansel:1999}
K.~M. Ansel, L.~J. McHeyzer-Williams, V.~N. Ngo, M.~G. McHeyzer-Williams, and
  J.~G. Cyster (1999).
\newblock {I}n vivo-activated {CD}4 {T} cells upregulate {CXC} chemokine
  receptor 5 and reprogram their response to lymphoid chemokines.
\newblock {\em J.~Exp.~Med.\/}, {\bf 190}(8):1123--34.

\bibitem{Fillatreau:2003}
S.~Fillatreau and D.~Gray (2003).
\newblock {T} cell accumulation in {B} cell follicles is regulated by dendritic
  cells and is independent of {B} cell activation.
\newblock {\em J.~Exp.~Med.\/}, {\bf 197}(2):195--206.

\bibitem{Miller:2002}
M.~J. Miller, S.~H. Wei, I.~Parker, and M.~D. Cahalan (2002).
\newblock {T}wo-photon imaging of lymphocyte motility and antigen response in
  intact lymph node.
\newblock {\em Science\/}, {\bf 296}(5574):1869--73.

\bibitem{Miller:2003}
M.~J. Miller, S.~H. Wei, M.~D. Cahalan, and I.~Parker (2003).
\newblock {A}utonomous {T} cell trafficking examined in vivo with intravital
  two-photon microscopy.
\newblock {\em Proc.~Natl.~Acad.~Sci.~U.S.A.\/}, {\bf 100}(5):2604--9.

\bibitem{Okada:2002}
T.~Okada, V.~N. Ngo, E.~H. Ekland, R.~Forster, M.~Lipp, D.~R. Littman, and
  J.~G. Cyster (2002).
\newblock {C}hemokine requirements for {B} cell entry to lymph nodes and
  {P}eyer's patches.
\newblock {\em J.~Exp.~Med.\/}, {\bf 196}(1):65--75.

\bibitem{Rangel-Moreno:2005}
J.~Rangel-Moreno, J.~Moyron-Quiroz, K.~Kusser, L.~Hartson, H.~Nakano, and T.~D.
  Randall (2005).
\newblock {R}ole of {CXC} chemokine ligand 13, {CC} chemokine ligand ({CCL})
  19, and {CCL}21 in the organization and function of nasal-associated lymphoid
  tissue.
\newblock {\em J.~Immunol.\/}, {\bf 175}(8):4904--13.

\bibitem{Luther:2002}
S.~A. Luther, A.~Bidgol, D.~C. Hargreaves, A.~Schmidt, Y.~Xu, J.~Paniyadi,
  M.~Matloubian, and J.~G. Cyster (2002).
\newblock {D}iffering activities of homeostatic chemokines {CCL}19, {CCL}21,
  and {CXCL}12 in lymphocyte and dendritic cell recruitment and lymphoid
  neogenesis.
\newblock {\em J.~Immunol.\/}, {\bf 169}(1):424--33.

\bibitem{Legler:1998}
D.~F. Legler, M.~Loetscher, R.~S. Roos, I.~Clark-Lewis, M.~Baggiolini, and
  B.~Moser (1998).
\newblock {B} cell-attracting chemokine 1, a human {CXC} chemokine expressed in
  lymphoid tissues, selectively attracts {B} lymphocytes via {BLR}1/{CXCR}5.
\newblock {\em J.~Exp.~Med.\/}, {\bf 187}(4):655--60.

\bibitem{Gunn:1998}
M.~D. Gunn, V.~N. Ngo, K.~M. Ansel, E.~H. Ekland, J.~G. Cyster, and L.~T.
  Williams (1998).
\newblock {A} {B}-cell-homing chemokine made in lymphoid follicles activates
  {B}urkitt's lymphoma receptor-1.
\newblock {\em Nature\/}, {\bf 391}(6669):799--803.

\bibitem{Gommerman:2002}
J.~L. Gommerman, F.~Mackay, E.~Donskoy, W.~Meier, P.~Martin, and J.~L. Browning
  (2002).
\newblock {M}anipulation of lymphoid microenvironments in nonhuman primates by
  an inhibitor of the lymphotoxin pathway.
\newblock {\em J.~Clin.~Invest.\/}, {\bf 110}(9):1359--69.

\bibitem{Huber:2005}
C.~Huber, C.~Thielen, H.~Seeger, P.~Schwarz, F.~Montrasio, M.~R. Wilson,
  E.~Heinen, Y.~X. Fu, G.~Miele, and A.~Aguzzi (2005).
\newblock {L}ymphotoxin-beta receptor-dependent genes in lymph node and
  follicular dendritic cell transcriptomes.
\newblock {\em J.~Immunol.\/}, {\bf 174}(9):5526--36.

\bibitem{Ohtsuka:1992}
A.~Ohtsuka, A.~J. Piazza, T.~H. Ermak, and R.~L. Owen (1992).
\newblock {C}orrelation of extracellular matrix components with the
  cytoarchitecture of mouse {P}eyer's patches.
\newblock {\em Cell Tissue Res.\/}, {\bf 269}(3):403--10.

\bibitem{Castanos-velez:1995}
E.~Castanos-Velez, P.~Biberfeld, and M.~Patarroyo (1995).
\newblock {E}xtracellular matrix proteins and integrin receptors in reactive
  and non-reactive lymph nodes.
\newblock {\em Immunology\/}, {\bf 86}(2):270--8.

\bibitem{Ushiki:1995}
T.~Ushiki, O.~Ohtani, and K.~Abe (1995).
\newblock {S}canning electron microscopic studies of reticular framework in the
  rat mesenteric lymph node.
\newblock {\em Anat.~Rec.\/}, {\bf 241}(1):113--22.

\bibitem{Katakai:2004}
T.~Katakai, T.~Hara, M.~Sugai, H.~Gonda, and A.~Shimizu (2004).
\newblock {L}ymph node fibroblastic reticular cells construct the stromal
  reticulum via contact with lymphocytes.
\newblock {\em J.~Exp.~Med.\/}, {\bf 200}(6):783--95.

\bibitem{Satoh:1997}
T.~Satoh, R.~Takeda, H.~Oikawa, and R.~Satodate (1997).
\newblock {I}mmunohistochemical and structural characteristics of the reticular
  framework of the white pulp and marginal zone in the human spleen.
\newblock {\em Anat.~Rec.\/}, {\bf 249}(4):486--94.

\bibitem{Liakka:1992}
K.~A. Liakka and H.~I. Autio-Harmainen (1992).
\newblock {D}istribution of the extracellular matrix proteins tenascin,
  fibronectin, and vitronectin in fetal, infant, and adult human spleens.
\newblock {\em J.~Histochem.~Cytochem.\/}, {\bf 40}(8):1203--10.

\bibitem{Maatta:2004}
M.~Maatta, S.~Salo, K.~Tasanen, Y.~Soini, A.~Liakka, L.~Bruckner-Tuderman, and
  H.~Autio-Harmainen (2004).
\newblock {D}istribution of basement membrane anchoring molecules in normal and
  transformed endometrium: altered expression of laminin gamma2 chain and
  collagen type {XVII} in endometrial adenocarcinomas.
\newblock {\em J.~Mol.~Histol.\/}, {\bf 35}(8-9):715--22.

\bibitem{Braun:2004}
A.~Braun, S.~Takemura, A.~N. Vallejo, J.~J. Goronzy, and C.~M. Weyand (2004).
\newblock {L}ymphotoxin beta-mediated stimulation of synoviocytes in rheumatoid
  arthritis.
\newblock {\em Arthritis Rheum.\/}, {\bf 50}(7):2140--50.

\bibitem{Kerjaschki:2005}
D.~Kerjaschki (2005).
\newblock {T}he crucial role of macrophages in lymphangiogenesis.
\newblock {\em J.~Clin.~Invest.\/}, {\bf 115}(9):2316--9.

\bibitem{Hamrah:2003}
P.~Hamrah, L.~Chen, Q.~Zhang, and M.~R. Dana (2003).
\newblock {N}ovel expression of vascular endothelial growth factor receptor
  ({VEGFR})-3 and {VEGF}-c on corneal dendritic cells.
\newblock {\em Am.~J.~Pathol.\/}, {\bf 163}(1):57--68.

\bibitem{Aurora:2005}
A.~B. Aurora, P.~Baluk, D.~Zhang, S.~S. Sidhu, G.~M. Dolganov, C.~Basbaum,
  D.~M. McDonald, and N.~Killeen (2005).
\newblock {I}mmune complex-dependent remodeling of the airway vasculature in
  response to a chronic bacterial infection.
\newblock {\em J.~Immunol.\/}, {\bf 175}(10):6319--26.

\bibitem{Ruddell:2003}
A.~Ruddell, P.~Mezquita, K.~A. Brandvold, A.~Farr, and B.~M. Iritani (2003).
\newblock {B} lymphocyte-specific c-{M}yc expression stimulates early and
  functional expansion of the vasculature and lymphatics during
  lymphomagenesis.
\newblock {\em Am.~J.~Pathol.\/}, {\bf 163}(6):2233--45.

\bibitem{Frater-Schroder:1987}
M.~Frater-Schroder, W.~Risau, R.~Hallmann, P.~Gautschi, and P.~Bohlen (1987).
\newblock Tumor necrosis factor type $\alpha$, a potent inhibitor of
  endothelial cell growth in vitro, is angiogenic in vivo.
\newblock {\em Proc.~Natl.~Acad.~Sci.~U.S.A.\/}, {\bf 84}(15):5277--81.

\bibitem{Corcione:1997}
A.~Corcione, L.~Ottonello, G.~Tortolina, P.~Tasso, F.~Ghiotto, I.~Airoldi,
  G.~Taborelli, F.~Malavasi, F.~Dallegri, and V.~Pistoia (1997).
\newblock {R}ecombinant tumor necrosis factor enhances the locomotion of memory
  and naive {B} lymphocytes from human tonsils through the selective engagement
  of the type {II} receptor.
\newblock {\em Blood\/}, {\bf 90}(11):4493--501.

\bibitem{Patterson:1996}
C.~Patterson, M.~A. Perrella, W.~O. Endege, M.~Yoshizumi, M.~E. Lee, and
  E.~Haber (1996).
\newblock Downregulation of vascular endothelial growth factor receptors by
  tumor necrosis factor-alpha in cultured human vascular endothelial cells.
\newblock {\em J.~Clin.~Invest.\/}, {\bf 98}(2):490--6.

\bibitem{Yoshida:1999}
H.~Yoshida, K.~Honda, R.~Shinkura, S.~Adachi, S.~Nishikawa, K.~Maki, K.~Ikuta,
  and S.~I. Nishikawa (1999).
\newblock {IL}-7 receptor $\alpha$+ {CD}3(-) cells in the embryonic intestine
  induces the organizing center of {P}eyer's patches.
\newblock {\em Int.~Immunol.\/}, {\bf 11}(5):643--55.

\bibitem{Hashi:2001}
H.~Hashi, H.~Yoshida, K.~Honda, S.~Fraser, H.~Kubo, M.~Awane, A.~Takabayashi,
  H.~Nakano, Y.~Yamaoka, and S.~Nishikawa (2001).
\newblock {C}ompartmentalization of {P}eyer's patch anlagen before lymphocyte
  entry.
\newblock {\em J.~Immunol.\/}, {\bf 166}(6):3702--9.

\bibitem{Kim:2000}
D.~Kim, R.~E. Mebius, J.~D. MacMicking, S.~Jung, T.~Cupedo, Y.~Castellanos,
  J.~Rho, B.~R. Wong, R.~Josien, N.~Kim, P.~D. Rennert, and Y.~Choi (2000).
\newblock {R}egulation of peripheral lymph node genesis by the tumor necrosis
  factor family member {TRANCE}.
\newblock {\em J.~Exp.~Med.\/}, {\bf 192}(10):1467--78.

\bibitem{Cupedo:2004}
T.~Cupedo, M.~F. Vondenhoff, E.~J. Heeregrave, A.~E. De~Weerd, W.~Jansen, D.~G.
  Jackson, G.~Kraal, and R.~E. Mebius (2004).
\newblock {P}resumptive lymph node organizers are differentially represented in
  developing mesenteric and peripheral nodes.
\newblock {\em J.~Immunol.\/}, {\bf 173}(5):2968--75.

\bibitem{Balogh:2001}
P.~Balogh, Y.~Aydar, J.~G. Tew, and A.~K. Szakal (2001).
\newblock {O}ntogeny of the follicular dendritic cell phenotype and function in
  the postnatal murine spleen.
\newblock {\em Cell.~Immunol.\/}, {\bf 214}(1):45--53.

\bibitem{Fu:1997}
Y.~X. Fu, H.~Molina, M.~Matsumoto, G.~Huang, J.~Min, and D.~D. Chaplin (1997).
\newblock {L}ymphotoxin-alpha ({LT}alpha) supports development of splenic
  follicular structure that is required for {I}g{G} responses.
\newblock {\em J.~Exp.~Med.\/}, {\bf 185}(12):2111--20.

\bibitem{Fu:1998}
Y.~X. Fu, G.~Huang, Y.~Wang, and D.~D. Chaplin (1998).
\newblock {B} lymphocytes induce the formation of follicular dendritic cell
  clusters in a lymphotoxin $\alpha$-dependent fashion.
\newblock {\em J.~Exp.~Med.\/}, {\bf 187}(7):1009--18.

\bibitem{Odell:1981}
G.~M. Odell, G.~Oster, P.~Alberch, and B.~Burnside (1981).
\newblock {T}he mechanical basis of morphogenesis. {I}. {E}pithelial folding
  and invagination.
\newblock {\em Dev.~Biol.\/}, {\bf 85}(2):446--62.

\bibitem{Palsson:2000}
E.~Palsson and H.~G. Othmer (2000).
\newblock {A} model for individual and collective cell movement in
  {D}ictyostelium discoideum.
\newblock {\em Proc.~Natl.~Acad.~Sci.~U.S.A.\/}, {\bf 97}(19):10448--53.

\bibitem{Dallon:2004}
J.~C. Dallon and H.~G. Othmer (2004).
\newblock {H}ow cellular movement determines the collective force generated by
  the {D}ictyostelium discoideum slug.
\newblock {\em J.~Theor.~Biol.\/}, {\bf 231}(2):203--22.

\bibitem{Drasdo:1995}
D.~Drasdo, R.~Kree, and J.~S. McCaskill (1995).
\newblock Monte carlo approach to tissue-cell populations.
\newblock {\em Phys.~Rev.~E\/}, {\bf 52}(6):6635--57.

\bibitem{Drasdo:2004}
D.~Drasdo, S.~D.~S. H{\"o}hme, and A.~Deutsch (2004).
\newblock {\em {C}ell-based models of avascular tumor growth.\/}, pages
  367--378.
\newblock Birkh{\"a}user, Basel.

\bibitem{Honda:2004}
H.~Honda, M.~Tanemura, and T.~Nagai (2004).
\newblock {A} three-dimensional vertex dynamics cell model of space-filling
  polyhedra simulating cell behavior in a cell aggregate.
\newblock {\em J.~Theor.~Biol.\/}, {\bf 226}(4):439--53.

\bibitem{Gunzer:2004}
M.~Gunzer, C.~Weishaupt, A.~Hillmer, Y.~Basoglu, P.~Friedl, K.~E. Dittmar,
  W.~Kolanus, G.~Varga, and S.~Grabbe (2004).
\newblock {A} spectrum of biophysical interaction modes between {T} cells and
  different antigen-presenting cells during priming in 3-{D} collagen and in
  vivo.
\newblock {\em Blood\/}, {\bf 104}(9):2801--9.

\bibitem{mehe:2005}
M.~E. Meyer-Hermann and P.~K. Maini (2005).
\newblock Interpreting two-photon imaging data of lymphocyte motility.
\newblock {\em Phys.~Rev.~E\/}, {\bf 71}(6 Pt 1):061912.

\bibitem{Albrecht:1998}
E.~Albrecht and H.~R. Petty (1998).
\newblock {C}ellular memory: neutrophil orientation reverses during temporally
  decreasing chemoattractant concentrations.
\newblock {\em Proc.~Natl.~Acad.~Sci.~U.S.A.\/}, {\bf 95}(9):5039--44.

\bibitem{Ehrengruber:1996}
M.~U. Ehrengruber, D.~A. Deranleau, and T.~D. Coates (1996).
\newblock {S}hape oscillations of human neutrophil leukocytes: characterization
  and relationship to cell motility.
\newblock {\em J.~Exp.~Biol.\/}, {\bf 199}(4):741--7.

\bibitem{Meineke:2001}
F.~A. Meineke, C.~S. Potten, and M.~Loeffler (2001).
\newblock {C}ell migration and organization in the intestinal crypt using a
  lattice-free model.
\newblock {\em Cell Prolif.\/}, {\bf 34}(4):253--66.

\bibitem{Johnson:1971}
K.~Johnson, K.~Kendall, and A.~Roberts (1971).
\newblock Surface energy and the contact of elastic solids.
\newblock {\em Proc.~R.~Soc.~London A\/}, {\bf 324}(1558):303--13.

\bibitem{Schnell:2004}
S.~Schnell and T.~E. Turner (2004).
\newblock Reaction kinetics in intracellular environments with macromolecular
  crowding: simulations and rate laws.
\newblock {\em Prog.~Biophys.~Mol.~Biol.\/}, {\bf 85}(2-3):235--60.

\bibitem{Thompson:1984}
C.~B. Thompson, I.~Scher, M.~E. Schaefer, T.~Lindsten, F.~D. Finkelman, and
  J.~J. Mond (1984).
\newblock {S}ize-dependent {B} lymphocyte subpopulations: relationship of cell
  volume to surface phenotype, cell cycle, proliferative response, and
  requirements for antibody production to {TNP}-{F}icoll and {TNP}-{BA}.
\newblock {\em J.~Immunol.\/}, {\bf 133}(5):2333--42.

\bibitem{Bornens:1989}
M.~Bornens, M.~Paintrand, and C.~Celati (1989).
\newblock {T}he cortical microfilament system of lymphoblasts displays a
  periodic oscillatory activity in the absence of microtubules: implications
  for cell polarity.
\newblock {\em J.~Cell Biol.\/}, {\bf 109}(3):1071--83.

\bibitem{Haston:1984}
W.~S. Haston and J.~M. Shields (1984).
\newblock {C}ontraction waves in lymphocyte locomotion.
\newblock {\em J.~Cell Sci.\/}, {\bf 68}:227--41.

\bibitem{Bausch:1998}
A.~R. Bausch, F.~Ziemann, A.~A. Boulbitch, K.~Jacobson, and E.~Sackmann (1998).
\newblock {L}ocal measurements of viscoelastic parameters of adherent cell
  surfaces by magnetic bead microrheometry.
\newblock {\em Biophys.~J.\/}, {\bf 75}(4):2038--49.

\bibitem{Bausch:1999}
A.~R. Bausch, W.~Moller, and E.~Sackmann (1999).
\newblock {M}easurement of local viscoelasticity and forces in living cells by
  magnetic tweezers.
\newblock {\em Biophys.~J.\/}, {\bf 76}(1 Pt 1):573--9.

\bibitem{Forgacs:1998}
G.~Forgacs, R.~A. Foty, Y.~Shafrir, and M.~S. Steinberg (1998).
\newblock {V}iscoelastic properties of living embryonic tissues: a quantitative
  study.
\newblock {\em Biophys.~J.\/}, {\bf 74}(5):2227--34.

\bibitem{Chu:2005}
Y.~S. Chu, S.~Dufour, J.~P. Thiery, E.~Perez, and F.~Pincet (2005).
\newblock {J}ohnson-{K}endall-{R}oberts theory applied to living cells.
\newblock {\em Phys.~Rev.~Lett.\/}, {\bf 94}(2):028102.

\bibitem{Maniotis:1997}
A.~J. Maniotis, C.~S. Chen, and D.~E. Ingber (1997).
\newblock {D}emonstration of mechanical connections between integrins,
  cytoskeletal filaments, and nucleoplasm that stabilize nuclear structure.
\newblock {\em Proc.~Natl.~Acad.~Sci.~U.S.A.\/}, {\bf 94}(3):849--54.

\bibitem{Hategan:2003}
A.~Hategan, R.~Law, S.~Kahn, and D.~E. Discher (2003).
\newblock {A}dhesively-tensed cell membranes: lysis kinetics and atomic force
  microscopy probing.
\newblock {\em Biophys.~J.\/}, {\bf 85}(4):2746--59.

\bibitem{Moy:1999}
V.~T. Moy, Y.~Jiao, T.~Hillmann, H.~Lehmann, and T.~Sano (1999).
\newblock Adhesion energy of receptor-mediated interaction measured by elastic
  deformation.
\newblock {\em Biophys.~J.\/}, {\bf 76}(3):1632--8.

\bibitem{Verdier:2003}
C.~Verdier (2003).
\newblock {R}heological properties of living materials. {F}rom cells to
  tissues.
\newblock {\em J.~Theo.~Med.\/}, {\bf 5}(2):67--91.

\bibitem{Balaban:2001}
N.~Q. Balaban, U.~S. Schwarz, D.~Riveline, P.~Goichberg, G.~Tzur, I.~Sabanay,
  D.~Mahalu, S.~Safran, A.~Bershadsky, L.~Addadi, and B.~Geiger (2001).
\newblock {F}orce and focal adhesion assembly: a close relationship studied
  using elastic micropatterned substrates.
\newblock {\em Nature Cell Biol.\/}, {\bf 3}:466--472.

\bibitem{Burton:1999}
K.~Burton, J.~H. Park, and D.~L. Taylor (1999).
\newblock {K}eratocytes generate traction forces in two phases.
\newblock {\em Mol.~Biol.~Cell\/}, {\bf 10}(11):3745--69.

\bibitem{Schwarz:2002b}
U.~S. Schwarz, N.~Q. Balaban, D.~Riveline, A.~Bershadsky, B.~Geiger, and S.~A.
  Safran (2002).
\newblock Calculation of forces at focal adhesions from elastic substrate data:
  the effect of localized force and the need for regularization.
\newblock {\em Biophys.~J.\/}, {\bf 83}(3):1380--94.

\bibitem{Galbraith:1999}
C.~G. Galbraith and M.~P. Sheetz (1999).
\newblock {K}eratocytes pull with similar forces on their dorsal and ventral
  surfaces.
\newblock {\em J.~Cell.~Biol.\/}, {\bf 147}(6):1313--24.

\bibitem{Lauffenburger:1996}
D.~A. Lauffenburger and A.~F. Horwitz (1996).
\newblock {C}ell migration: a physically integrated molecular process.
\newblock {\em Cell\/}, {\bf 84}(3):359--69.

\bibitem{Randolph:2005}
G.~J. Randolph, V.~Angeli, and M.~A. Swartz (2005).
\newblock {D}endritic-cell trafficking to lymph nodes through lymphatic
  vessels.
\newblock {\em Nat.~Rev.~Immunol.\/}, {\bf 5}(8):617--28.

\bibitem{Sacca:1998}
R.~Sacca, C.~A. Cuff, W.~Lesslauer, and N.~H. Ruddle (1998).
\newblock {D}ifferential activities of secreted lymphotoxin-alpha3 and membrane
  lymphotoxin-alpha1beta2 in lymphotoxin-induced inflammation: critical role of
  {TNF} receptor 1 signaling.
\newblock {\em J.~Immunol.\/}, {\bf 160}(1):485--91.

\bibitem{Lin:2004}
F.~Lin, C.~M. Nguyen, S.~J. Wang, W.~Saadi, S.~P. Gross, and N.~L. Jeon (2004).
\newblock {E}ffective neutrophil chemotaxis is strongly influenced by mean
  {IL}-8 concentration.
\newblock {\em Biochem.~Biophys.~Res.~Commun.\/}, {\bf 319}(2):576--81.

\bibitem{Pelletier:2000}
A.~J. Pelletier, L.~J. van~der Laan, P.~Hildbrand, M.~A. Siani, D.~A. Thompson,
  P.~E. Dawson, B.~E. Torbett, and D.~R. Salomon (2000).
\newblock {P}resentation of chemokine {SDF}-1 alpha by fibronectin mediates
  directed migration of {T} cells.
\newblock {\em Blood\/}, {\bf 96}(8):2682--90.

\bibitem{Slimani:2003}
H.~Slimani, N.~Charnaux, E.~Mbemba, L.~Saffar, R.~Vassy, C.~Vita, and
  L.~Gattegno (2003).
\newblock {B}inding of the {CC}-chemokine {RANTES} to syndecan-1 and syndecan-4
  expressed on {H}e{L}a cells.
\newblock {\em Glycobiology\/}, {\bf 13}(9):623--34.

\bibitem{Vissers:2001}
J.~L. Vissers, F.~C. Hartgers, E.~Lindhout, C.~G. Figdor, and G.~J. Adema
  (2001).
\newblock {BLC} ({CXCL}13) is expressed by different dendritic cell subsets in
  vitro and in vivo.
\newblock {\em Eur.~J.~Immunol.\/}, {\bf 31}(5):1544--9.

\bibitem{Hu:2004}
W.-S. Hu.
\newblock Stoichiometry and kinetics of cell growth and product formation.
\newline\urlprefix\url{http://hugroup.cems.umn.edu/Cell\_Technology/cd-rom/Stoichiometry\%20and\%20 Kinetics/Stoichiometry\%20and\%20Cell\%20Kinetics.pdf}

\end{thebibliography}

\end{document}